\documentclass[12pt]{iopart}
\usepackage{iopams}
\usepackage{setstack}
\usepackage{graphicx}
\usepackage{subfigure}

\usepackage{color}

\begin{document}

\title[The directed feedback vertex set problem]{A spin glass
  approach to the directed feedback vertex set problem}

\author{Hai-Jun Zhou}

\address{
 State Key Laboratory of Theoretical Physics,
 Institute of Theoretical Physics,
 Chinese Academy of Sciences, Zhong-Guan-Cun East Road 55,
 Beijing 100190, China
}
\ead{zhouhj@itp.ac.cn}
\vspace{10pt}
\begin{indented}
\item[]Janary 17, 2016 (first version); April 04, 2016 (revised version)
\end{indented}

\begin{abstract}
  A directed graph (digraph) is formed by vertices and arcs (directed edges)
  from one vertex to another. A feedback vertex set (FVS) is a set of vertices
  that contains at least one vertex of every directed cycle in this digraph.
  The directed feedback vertex set problem aims at constructing a FVS of
  minimum cardinality. This is a fundamental cycle-constrained hard
  combinatorial optimization problem with wide practical applications.
  In this paper we construct a spin glass model for the directed FVS problem
    by converting the global cycle constraints into local arc constraints,
    and study this model through the replica-symmetric (RS) mean field theory
    of statistical physics. We then implement a belief propagation-guided
    decimation (BPD) algorithm for single digraph instances.
    The BPD algorithm slightly outperforms the simulated annealing algorithm
    on large random graph instances. The predictions of the RS mean field
    theory are noticeably lower than the BPD results, possibly  due to its
    neglect of cycle-caused long range correlations.
\end{abstract}


\section{Introduction}

Directed graphs (digraphs) are widely used to describe interactions in
technological, biological, and social complex systems
\cite{He-Liu-Wang-2009,Albert-Barabasi-2002,Ermann-etal-2015}.
A digraph is composed of vertices and arcs (directed edges) which link
between vertices. For instance, the gene regulation system of a biological
cell can be modeled as a digraph of genes, in which an arc points from gene 
$A$ to gene $B$ if $A$ regulates the expression of $B$ 
\cite{Li-etal-2004,VelizCuba-etal-2014}.
Real-world complex systems are full of feedback interactions and adaptation
mechanisms, and the digraphs of these systems usually contain an
abundant number of directed cycles, which make the system's dynamical
properties difficult to predict and to externally control
\cite{Sorrentino-2007,Liu-Slotine-Barabasi-2011}.

Not all vertices and arcs are equally important in feedback interactions.
Some vertices and arcs may participate in a much greater number of
directed cycles than other vertices and arcs.
A minimum feedback vertex set (FVS) contains 
a smallest number of vertices whose deletion from the digraph destroys all
the directed cycles. Similarly, a minimum feedback arc set (FAS) is
an arc set of smallest cardinality such that every directed cycle of the
digraph has at least one arc in this set. 
In terms of collective effect, the vertices in a FVS and the arcs
in a FAS may be most significant to the dynamical complexity of a
complex networked system
\cite{Fiedler-etal-2013,Mochizuki-etal-2013,Festa-Pardalos-Resende-1999},
  and these vertices and arcs may also be optimal targets of distributed
  network attack processes \cite{Mugisha-Zhou-2016,Braunstein-etal-2016}.

The directed feedback problems, namely constructing a minimum FVS and
a minimum FAS for a generic digraph, are combinatorial optimization problems
in the nondeterministic polynomial-hard (NP-hard) class
and therefore are intrinsically very difficult \cite{Garey-Johnson-1979}.
Some progresses have been achieved by computer scientists and applied
mathematicians in understanding this classic hard problem since the
1990s \cite{Festa-Pardalos-Resende-1999}. Approximate algorithms with
provable bounds have been designed (see, for example
\cite{Even-etal-1998,Cai-Deng-Zang-2001,Chen-Liu-Lu-etal-2008})
and efficient heuristic algorithms such as greedy local search
\cite{Pardalos-Qian-Resende-1999} and simulated annealing
\cite{Galinier-Lemamou-Bouzidi-2013} have been implemented
and tested on benchmark small problem instances.
Yet unlike the FVS problem on an undirected graph
\cite{Bafna-Berman-Fujito-1999,Bau-Wormald-Zhou-2002},
it is still very difficult to give tight bounds on the
minimal cardinality of the directed FVS and FAS problems.

In this paper we study the directed FVS problem using statistical
physics methods.
  Our method is also applicable to the FAS problem since it
  is essentially equivalent to the FVS problem \cite{Even-etal-1998}.
We construct a spin glass model for the directed
FVS problem and derive the replica-symmetric (RS) mean field theory for
this model. The mean field theory enables us to estimate the
minimum directed FVS cardinality for random digraphs.
Based on this mean field theory we implement a
belief propagation-guided decimation (BPD) algorithm and apply this
message-passing algorithm to large random digraph instances.
The BPD algorithm slightly outperforms the simulated annealing
algorithm \cite{Galinier-Lemamou-Bouzidi-2013} in terms of the
FVS cardinality and the arc density of the remaining directed acyclic
graph (DAG), while the computing times of the two algorithms are comparable to
each other. 
Our work will be helpful for future investigations on the dynamical
complexity of various real-world networked systems
  and for further studies of targeted attacks on directed
  networks.
In this paper we also point out a major difficulty of the RS mean 
field theory in treating directed cycle-caused long range correlations. 

Directed and undirected cycles cause long-range correlations and frustrations
in dynamical systems and spin glass models. They are very important
network structural properties, but theoretical efforts directly treating them
as constraints are quite lacking.
Directed cycles are global properties of a digraph (each of which may
involve a lot of vertices and arcs).
Compared with problems with local constraints (such as the $K$-satisfiability
problem \cite{Monasson-Zecchina-1996,Mezard-etal-2002,Mezard-Zecchina-2002}),
optimization problems constrained by directed cycles are much more difficult 
to tackle theoretically.
The present paper is a continuation of our earlier report \cite{Zhou-2013} which
treated the undirected FVS problem successfully. We realized that the
directness of the cycle constraints makes the directed FVS problem much
harder than its undirected counterpart.
It is clear that further efforts are needed
to overcome the gap between the BPD results and the RS
predictions in Fig.~\ref{fig:Rdfvs}. 

The next section defines the directed FVS and FAS problems and explains their
essential equivalence.
We then introduce a FVS spin glass model in Sec.~\ref{sec:model} and
study it by RS mean field theory in
Sec.~\ref{sec:RS}. In Sec.~\ref{sec:BPD} we describe the
BPD message-passing algorithm and compare its performance with the
simulated annealing algorithm on random digraphs.
We conclude this paper in Sec.~\ref{sec:conclusion}.
  The two appendices
  contain some additional technical discussions.

\section{The directed FVS and FAS problems}
\label{sec:FVS}

Consider a simple digraph  $G$. This digraph contains $N$ vertices whose
indices (say $i, j, k, \ldots$) range from integer values $1$ to $N$.
There are $M$ arcs in this digraph, each of which points from one vertex 
(say $i$) to another different vertex (say $j$) and is denoted as $[i, j]$.
There is no self-arc pointing from a vertex to itself, and
there is at most one arc  from any vertex $i$ to any
another different vertex $j$. The arc density $\alpha$ of a
digraph is just the ratio between arc number $M$ and
vertex number $N$, that is $\alpha \equiv \frac{M}{N}$.

Each vertex $i$ of the digraph has a positive weight $w_i$ whose
meaning is context-dependent.
For instance, in the problem of complex-system control
\cite{Liu-Slotine-Barabasi-2011}, the weight $w_i$ may be the
cost of monitoring the vertex $i$. The vertex weights
are fixed parameters of the digraph and  can not be modified.
In the actual numerical computations of this paper, each vertex weight
$w_i$ is set to be $w_i=1$ just for simplicity.

\begin{figure}
  \begin{center}
    \subfigure[]{
      \label{fig:digraph:a}
      \includegraphics[width=0.3\textwidth]{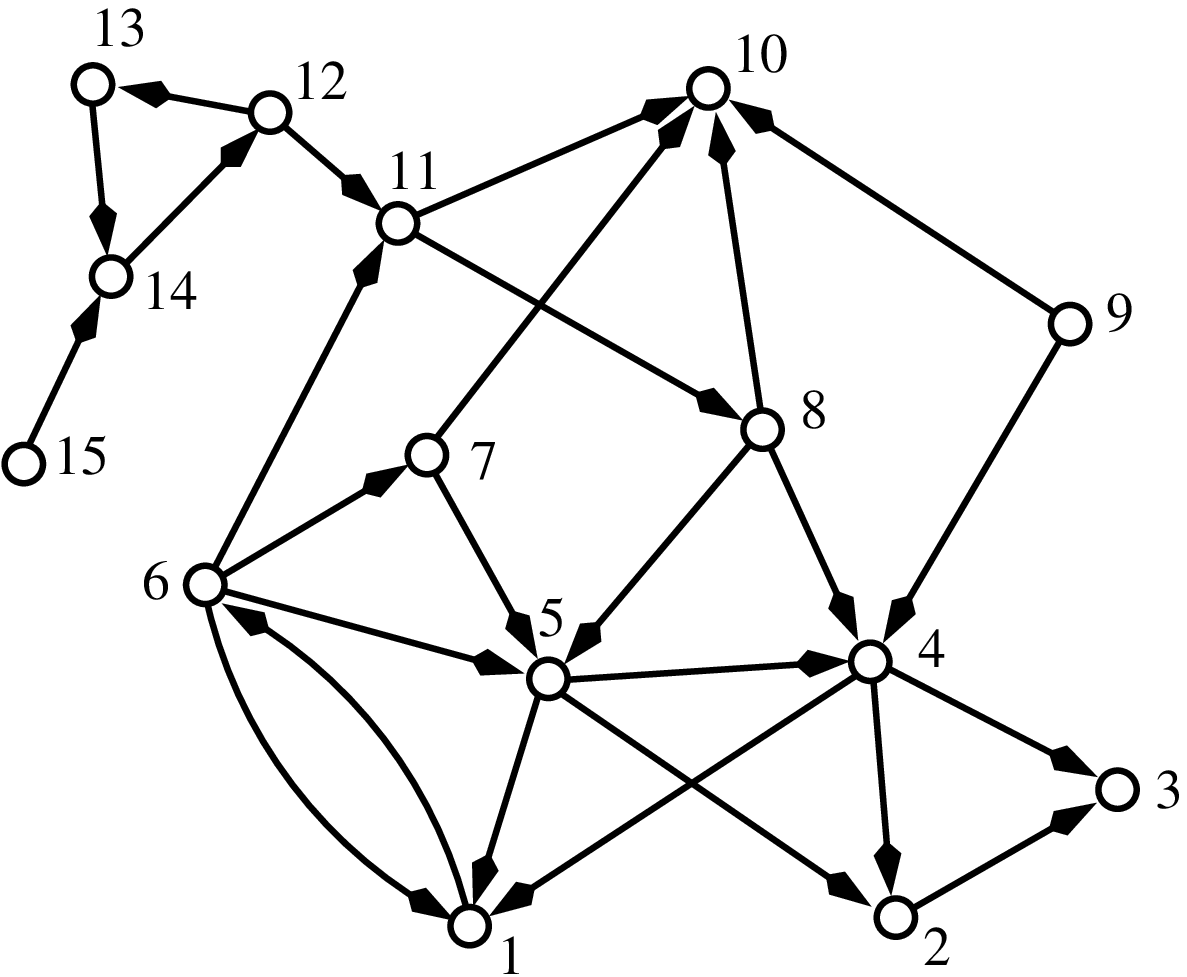}
    }
    \subfigure[]{
      \label{fig:digraph:b}
      \includegraphics[width=0.3\textwidth]{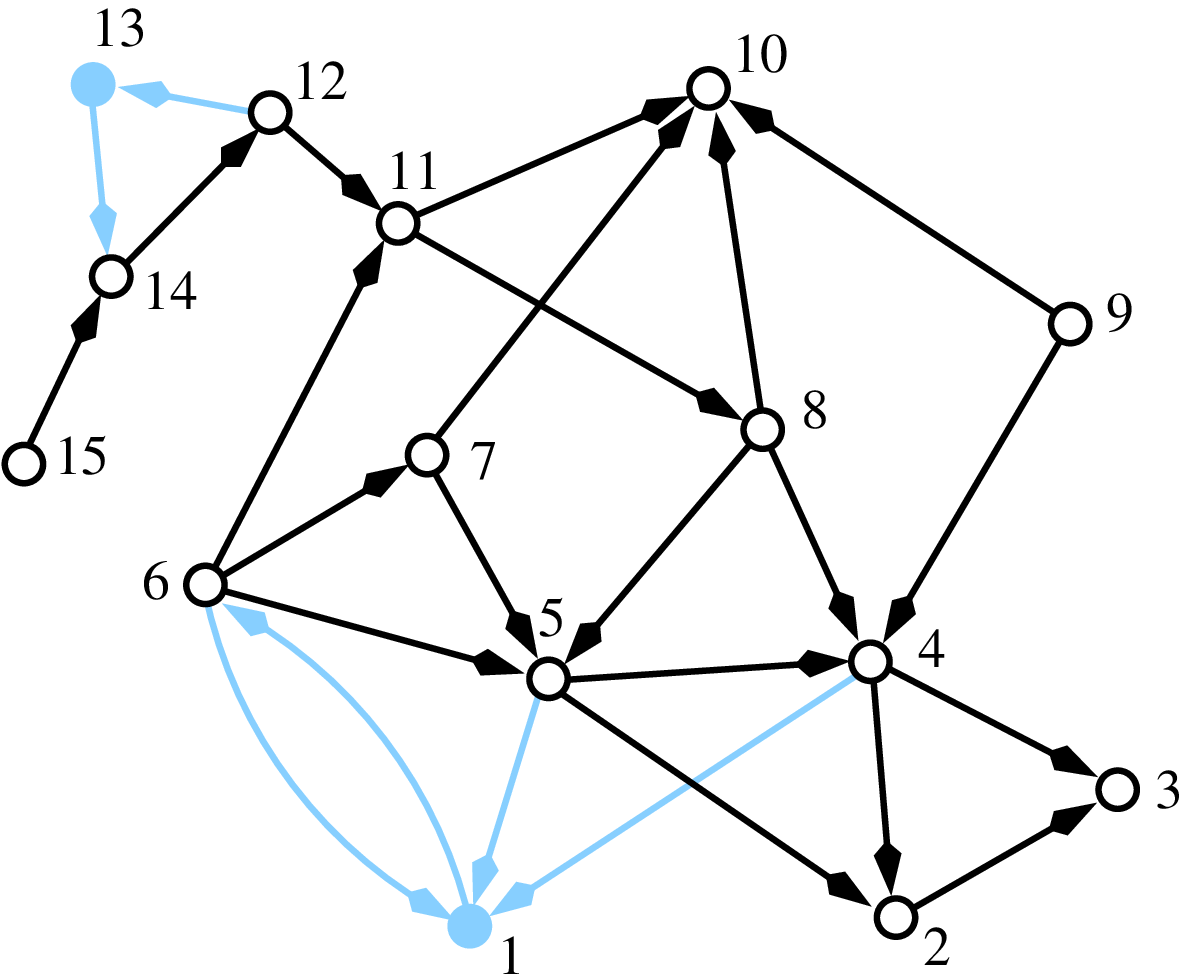}
    }
    \subfigure[]{
      \label{fig:digraph:c}
      \includegraphics[width=0.3\textwidth]{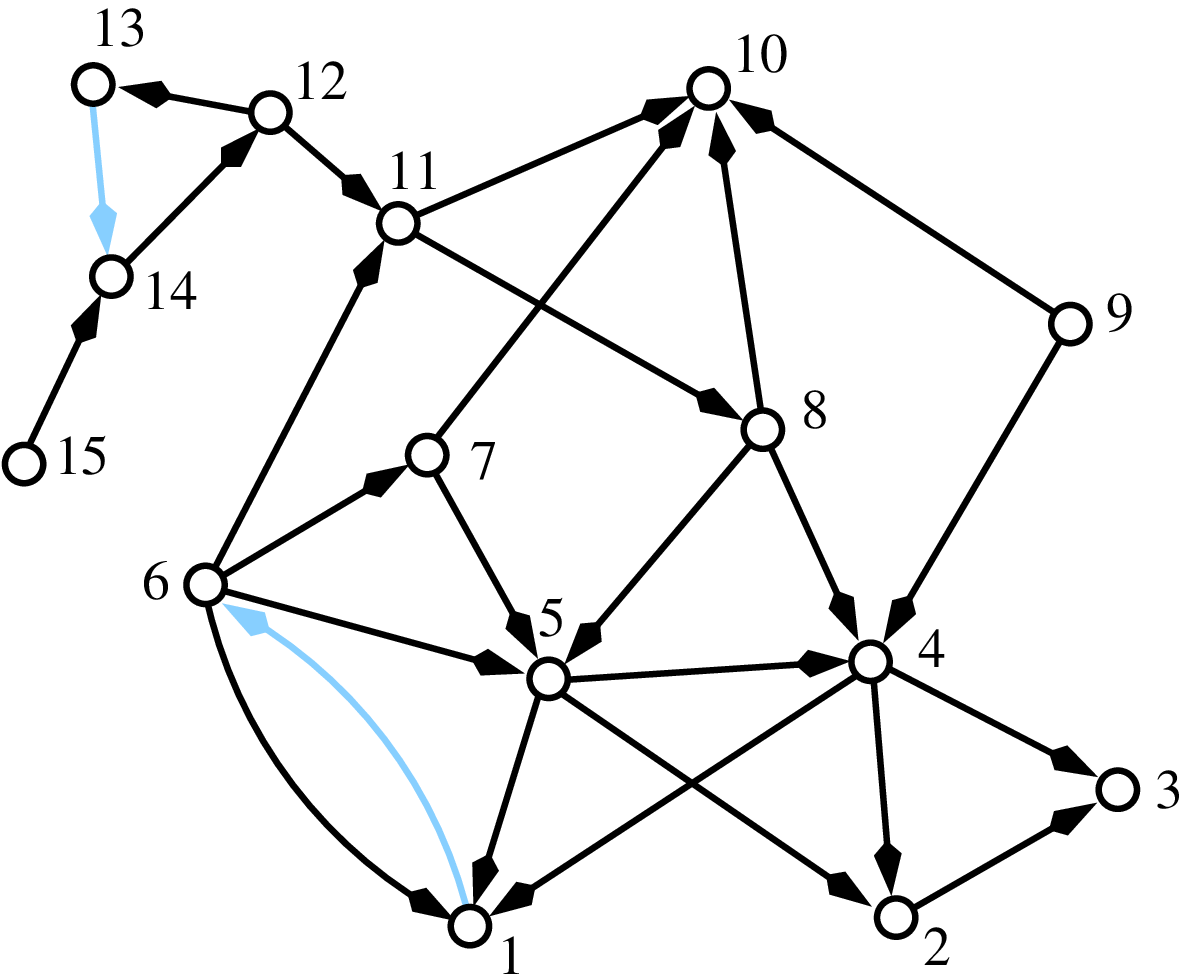}
    }
  \end{center}
  \caption{
    \label{fig:digraph}
    (a) A small digraph containing $N=15$ vertices and $M=25$ arcs, 
    (b) a minimum feedback vertex set $\{1, 13\}$ and
    (c) a minimum feedback arc set $\{[1,6], [13,14]\}$ for this digraph.
  }
\end{figure}

If there is an arc $[i, j]$ from a vertex
$i$ to another vertex $j$ but the reverse arc $[j, i]$ is absent,
then $j$ is said to be a \emph{child} of
$i$ and $i$ is said to be a \emph{parent} of $j$.
If there are two arcs $[i, j]$ and $[j, i]$ of reverse direction
between vertices $i$ and $j$, then
vertex $i$ is said to be a \emph{brother} of
vertex $j$ and $j$ a brother of $i$.
The in-degree and out-degree of vertex $j$ are defined as the total number of
arcs pointing to $j$ and pointing from $j$, respectively.
For example, vertex $5$ in Fig.~\ref{fig:digraph} is a
parent of vertex $1$ and is a child of vertex $6$, while
vertices $6$ and $1$ are brother of each other.

A directed path in a digraph $G$ is a sequence of arcs
which connect a
sequence of vertices, for example a directed path
$[i, j_1] , \ [j_1,  j_2] , \ \ldots, \ [j_{n-1} ,  j_{n}] , \ [j_n , j]$
from a start vertex $i$ to an end vertex $j$.
If the start vertex and the end vertex of a directed path are
identical, then this path is referred to as a directed cycle.
Some examples of directed paths and cycles can be found in the
small graph of Fig.~\ref{fig:digraph}.

A feedback vertex set of digraph $G$ is a subset
$\Gamma$ of the $N$ vertices
such that if all the vertices of this set and the attached arcs
are removed from $G$ the remaining digraph will have no directed cycle.
In other words, the set $\Gamma$ contains at least one vertex of
every directed cycle.
As an example, the set $\Gamma = \{1, 13\}$ is a FVS for the
small digraph of Fig.~\ref{fig:digraph}. Similarly, a feedback arc set 
is a subset of the $M$ arcs such that it contains at least
one arc of every directed cycle of digraph $G$. For the example of 
Fig.~\ref{fig:digraph} one can easily verify that the arc set
$\{[1, 6], [13, 14]\}$ is a minimum FAS.

The weight $w(\Gamma)$ of a feedback vertex set $\Gamma$ is just the
total weight of its ingredient vertices:
$w(\Gamma) \equiv \sum_{i\in \Gamma} w_i$.
The fundamental goal of the feedback vertex set problem is to construct
a FVS for a given digraph with total weight as small as
possible. This NP-hard problem has been studied by mathematicians
and computer scientists, see for example references
\cite{Festa-Pardalos-Resende-1999,Even-etal-1998,Cai-Deng-Zang-2001,Chen-Liu-Lu-etal-2008,Pardalos-Qian-Resende-1999,Galinier-Lemamou-Bouzidi-2013}, but
it appears that this important problem has not yet been treated by
statistical physics methods. Here we work on the directed FVS problem
using the mean field spin glass theory and the belief propagation
message-passing algorithm.

\begin{figure}
  \begin{center}
    \includegraphics[width=0.6\textwidth]{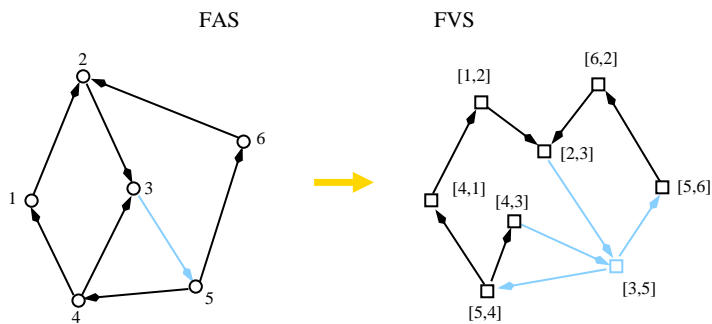}
  \end{center}
  \caption{
    \label{fig:fastofvs}
    The feedback arc set problem for a digraph $G$ (left) can be
    converted into a feedback vertex set problem for a new digraph $\tilde{G}$
    (right). Each
    arc $[i,j]$ of $G$ is mapped to a vertex (denoted by
    a square) in $\tilde{G}$, and an arc is created in
    $\tilde{G}$ from a vertex $[i, j]$ to another vertex $[k,l]$ 
    if and only if $j=k$. The left digraph becomes acyclic after arc $[3, 5]$
    is deleted, and correspondingly the right digraph becomes acyclic after 
    vertex $[3, 5]$ and its attached arcs are deleted.
  }
\end{figure}

The spin glass model of this work and the corresponding BPD algorithm is also
applicable to the FAS problem. This is because the FAS problem is equivalent
to the FVS problem
  on a modified digraph \cite{Even-etal-1998}
(Fig.~\ref{fig:fastofvs}). 
To construct a FAS for a given digraph $G$, we can first map each arc
$[i, j]$ of $G$ to a vertex (also denoted as $[i,j]$) of a new 
digraph $\tilde{G}$, and then we setup an arc in $\tilde{G}$
from a vertex $[i,j]$ to another
vertex $[k,l]$ if and only if $j=k$. It is easy to verify that there is
a one-to-one correspondence between a directed cycle of digraph $G$ and a 
directed cycle in the digraph $\tilde{G}$, and a FVS of digraph
$\tilde{G}$ is just a FAS of digraph $G$ (Fig.~\ref{fig:fastofvs}).

  In the present paper we focus on the directed FVS problem. Treatment of
  the FAS problem on random and real-world networks and its application to
  the network destruction problem will be reported in
  a separate paper \cite{Zhao-Zhou-2016}.

\section{Spin glass model}
\label{sec:model}

We now describe in detail a spin glass model for the directed FVS problem,
  which was briefly introduced in the last section of
  \cite{Zhou-2013} by the present author. Similar model systems have also
  been discussed by other authors, see, e.g., 
  \cite{Altarelli-Braunstein-DallAsta-Zecchina-2013,Altarelli-Braunstein-DallAsta-Zecchina-2013b,Guggiola-Semerjian-2015}.
First, a height state $h_i$ is assigned to each vertex $i$ of a digraph $G$.
This height state can take integer values in
the interval $0 \leq h_i \leq D$, with $D$ being the allowed maximal height
in the model.
The value of $D$ is adjustable in the theoretical calculations, but
it should not exceed the total number $N$ of vertices.
Setting $D$ to a larger value will result in better theoretical results.
On the other hand, the memory space and the computing time both
increase linearly with $D$. In this paper we mainly work with $D=200$ but
we will also discuss the effect of $D$ on the theoretical results.

A generic configuration of the system is denoted as $\underline{h}$ with
$\underline{h} \equiv \{h_1, h_2, \ldots, h_N\}$.
If the height of a vertex $i$ is $h_i = 0$, this vertex is referred to as
being unoccupied or empty, otherwise $i$ is referred to as being occupied.
To ensure that there are no directed cycles within the set of occupied
vertices, we impose the following constraint on each arc $[i, j]$ of digraph
$G$: if both vertex $i$ and vertex $j$ are occupied ($h_i >0$ and $h_j>0$),
then the height of vertex $i$ must be less than that of vertex $j$
($h_i < h_j$).  A configuration $\underline{h}$ is referred to as a
legal configuration if it satisfies all the $M$ arc constraints.

If $\underline{h}$ is a legal configuration of digraph $G$,
then the vertex heights along any directed path of occupied vertices
form a strictly increasing sequence. Therefore there must be no
directed cycles within the set of occupied vertices of configuration
$\underline{h}$ and the set of unoccupied vertices of this configuration
must form a FVS. In other words, each legal  configuration
$\underline{h}$ corresponds to exactly one FVS.

On the other hand, since two or more legal configurations may have
the identical set of unoccupied vertices, a single FVS
often corresponds to more than one legal configuration. In this
sense the above introduced arc constraints are not very restrictive.
We can achieve a one-to-one mapping between a legal configuration and a 
FVS by requiring that if the height of a vertex is positive it must exceed 
the maximal height of its parent and brother vertices by $1$
(see \ref{sec:modelB}).
But the spin glass model with such a set of more restrictive arc constraints 
is more complicate for numerical treatment,
  and it does not lead to better FVS solutions for single digraph instances.
  In the present paper we stick to the weaker arc constraints. A comparative
  study on this weaker model and the stronger model of \ref{sec:modelB} will
  be carried out in another technical paper.

For the convenience of theoretical development, if there is an arc
$[i, j]$ between a vertex $i$ and another vertex $j$ but the
reverse arc $[j, i]$ is absent, we denote this situation by
the notation $(i\rightarrow j)$; if both the forward arc $[i, j]$
and the reverse arc $[j, i]$ exist between vertices $i$ and $j$, we denote
this situation by the notation $(i \Leftrightarrow j)$. In later
discussions,  $(i\rightarrow j)$ and $(i\Leftrightarrow j)$ are
referred to as \emph{links} of digraph $G$.
A link factor is defined for each link $(i\rightarrow j)$ as
\begin{equation}
  \label{eq:linkc1}
  C_{(i\rightarrow j)}(h_i, h_j) = \delta_{h_j}^0 +
  \bigl(1-\delta_{h_j}^0 \bigr) \Theta(h_j - h_i) \; ,
\end{equation}
where $\delta_m^n$ is the Kronecker symbol such that $\delta_m^n=1$ if
$m=n$ and $\delta_m^n=0$ if $m\neq n$, and $\Theta(n)$ is the Heaviside
step function such that $\Theta(n) = 0$ for integers $n\leq 0$ and
$\Theta(n) = 1$ for $n \geq 1$.
The link factor $C_{(i\rightarrow j)} =1$ if either $h_j=0$ (vertex $j$ is
unoccupied) or $h_j > h_i$ (vertex $j$ is occupied and its height is
higher than that of vertex $i$), otherwise $C_{(i\rightarrow j)}=0$.
A different link factor is defined for each link $(i\Leftrightarrow j)$ as
\begin{equation}
  \label{eq:linkc2}
  C_{(i\Leftrightarrow j)}(h_i, h_j) = \delta_{h_i h_j}^0 \; .
\end{equation}
This link factor $C_{(i\Leftrightarrow j)}=1$ if either $h_i=0$ (vertex $i$ is
unoccupied) or $h_j=0$ (vertex $j$ is unoccupied) or both, otherwise its 
value is zero.

We now introduce a partition function $Z(x)$ for the digraph $G$ as
\cite{Zhou-2013}
\begin{equation}
  \hspace*{-1.8cm}
  Z(x) =  \sum\limits_{\underline{h}}
  \exp\Bigl[ x \sum\limits_{i=1}^{N} (1- \delta_{h_i}^{0} ) w_i \Bigr]
  \prod\limits_{(i\rightarrow j)\in G} C_{(i\rightarrow j)}(h_i, h_j)
  \prod\limits_{(k \Leftrightarrow l) \in G} C_{(k\Leftrightarrow l)}(h_k, h_l)  \; .
  \label{eq:Zxd1}
\end{equation}
Because of the product term of link factors in the above expression, only
legal configurations have positive contributions to the partition
function. The parameter $x$ of (\ref{eq:Zxd1}) is a re-weighting
parameter (the inverse temperature) which favors configurations of 
larger total weight of occupied vertices. As $x$ becomes sufficiently
large, the partition function $Z(x)$ will be contributed predominantly by
those height configurations $\underline{h}$ which correspond to minimum
feedback vertex sets.

Given a height configuration $\underline{h}$, the total number of occupied
vertices, $N_1(\underline{h})$, and
the total number of occupied arcs, $M_1(\underline{h})$, are
computed respectively through
\begin{equation}
  N_1(\underline{h})  =  \sum\limits_{i=1}^N
  \bigl( 1 - \delta_{h_i}^0 \bigr)\; , \quad \quad \quad
  M_1(\underline{h})  =  \sum\limits_{[i, j] \in G}
  \bigl(1- \delta_{h_i}^0
  \bigr) \bigl(1-\delta_{h_j}^0 \bigr) \; .
\end{equation}
The free entropy $\Phi(x)$ of the spin glass system is defined as
\begin{equation}
  \Phi(x) = \frac{1}{x} \ln Z(x) \; .
\end{equation}
For a digraph $G$ containing a large number $N$ of vertices, we expect
the free entropy $\Phi(x)$ to be an extensive thermodynamic quantity,
namely $\Phi(x) \simeq N \phi(x)$ with $\phi(x)$ being the free
entropy density. The free entropy density $\phi$ depends on the
the re-weighting parameter $x$ but not on the vertex number $N$ (for
$N$ sufficiently large).

An interesting connection which we wish to point out is that the directed FVS
problem can be viewed as an irreversible opinion propagation problem. 
Consider a simple contagion process in which an initially occupied vertex
will decay to be unoccupied if all its parent and brother vertices become 
unoccupied \cite{Chang-Lyuu-2012}.
For this process to reach all the vertices of the digraph
we need to fix some vertices to be unoccupied as the initial condition.
This set of externally fixed vertices must form a FVS. 
In this opinion-dynamics context the height $h_i$ of a vertex $i$ should
be understood as the discrete time at which vertex $i$ jumps from being
occupied to being unoccupied (see 
\cite{Altarelli-Braunstein-DallAsta-Zecchina-2013,Altarelli-Braunstein-DallAsta-Zecchina-2013b,Guggiola-Semerjian-2015,DelFerraro-Aurell-2015,Braunstein-etal-2016}
for related theoretical and computational efforts on irreversible and 
reversible opinion spreading processes).

\section{Replica-symmetric mean field theory}
\label{sec:RS}

We now study the model (\ref{eq:Zxd1}) by the RS mean field spin glass theory.
This theory is based on the Bethe-Peierls approximation
\cite{Mezard-Montanari-2009,Mezard-Parisi-2001,Bethe-1935}, and it can also
be derived through the method of partition function expansion
\cite{Chertkov-Chernyak-2006b,Zhou-Wang-2012,Xiao-Zhou-2011,Zhou-etal-2011}.

\subsection{Theory}

A quantity of central importance is the probability that a vertex belongs
to a FVS (i.e., being unoccupied). To calculate this marginal probability,
let us denote by $q_j^{h_j}$ the probability of vertex $j$ taking the
height state $h_j$. If there is a link between two vertices $i$ and $j$,
we denote by $q_{j\rightarrow i}^{h_j}$ the marginal probability that vertex
$j$ would take the height state $h_j$ if the effect of vertex $i$ to $j$
was not considered.

\begin{figure}
  \begin{center}
    \includegraphics[width=0.3\textwidth]{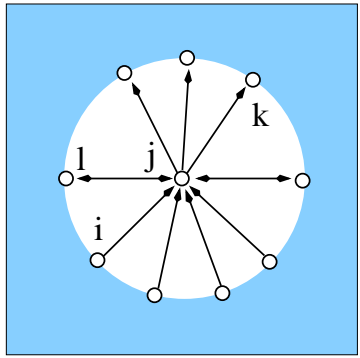}
  \end{center}
  \caption{
    \label{fig:BPapprox}
    The neighborhood of a vertex $j$. The whole square region represents the
    digraph $G$, and the shaded region represents the cavity digraph
    $G_{\backslash j}$ after vertex $j$ and all its attached arcs are removed
    from $G$. The neighboring vertices of $j$ can be divided into three
    disjoint sets, the parent set $p(j)$, the child set $c(j)$, and the
    brother set $b(j)$. For instance, vertex $i \in p(j)$,  $k\in c(j)$,
    and  $l\in b(j)$. Each double-arrowed link in this
    figure denotes two oppositely directed arcs.}
\end{figure}

Consider a randomly chosen vertex $j$ of the digraph $G$. In general
this vertex is connected to a set of parent, child, and brother
vertices (Fig.~\ref{fig:BPapprox}). Let
us denote by $p(j)$, $c(j)$, and $b(j)$ the set of parent, child, and
brother vertices of vertex $j$, respectively. In mathematical terms,
$p(j) \equiv  \{i : (i\rightarrow j)\in G\}$,
$c(j) \equiv  \{k : (j\rightarrow k)\in G\}$, 
and $b(j)  \equiv  \{l : (j\Leftrightarrow l) \in G\}$.
The set of neighboring vertices of vertex $j$, denoted as $\partial j$,
is the union of these three disjoint sets, namely 
$\partial j \equiv p(j) \cup c(j) \cup b(j)$. 

To compute the marginal probability $p_j^{h_j}$, let us first remove vertex
$j$ and all the attached arcs from digraph $G$. The remaining digraph is
referred to as a cavity graph and is denoted as $G_{\backslash j}$ 
(Fig.~\ref{fig:BPapprox}). As a first approximation we assume that in the
cavity digraph $G_{\backslash j}$ all the vertices in the set $\partial j$ are
mutually independent and therefore the joint distribution of these
vertices' height states can be written as a product of the marginal
height distributions of single vertices (the Bethe-Peierls approximation
\cite{Bethe-1935,Mezard-Montanari-2009}).
Under this approximation, when vertex $j$ is added to the cavity digraph
$G_{\backslash j}$ to form the whole digraph $G$, its marginal height
distribution $q_j^{h_j}$ can be expressed as
\begin{equation}
  q_j^{0} =
  \frac{1}{z_j} \; ,
  \quad \quad \quad 
  q_j^{h_j}
  = \frac{a_j(h_j)}{z_j}
  \quad (1 \leq h_j \leq D) \; ,
  \label{eq:qjmarginal}
\end{equation}
where $a_j(h)$ is a shorthand notation for the expression
\begin{equation}
  a_j(h) \equiv 
  e^{x w_j}
  \prod\limits_{l\in b(j)} q_{l\rightarrow j}^0
  \prod\limits_{i \in p(j)}
  \Bigl[
    \sum\limits_{h^\prime=0}^{h-1} q_{i\rightarrow j}^{h^\prime}
    \Bigr]
  \prod\limits_{k\in c(j)}
  \Bigl[
    q_{k \rightarrow j}^0 + \sum\limits_{h^{\prime\prime} \geq h+1}^{D}
    q_{k\rightarrow j}^{h^{\prime\prime}}
    \Bigr] \; ,
\end{equation}
and the normalization constant $z_j$ is expressed as
\begin{equation}
  z_j  \equiv 
  1 + e^{x w_j} \sum\limits_{h=1}^{D}
  \prod\limits_{l\in b(j)} q_{l\rightarrow j}^0
  \prod\limits_{i \in p(j)}
  \Bigl[
    \sum\limits_{h^\prime=0}^{h-1} q_{i\rightarrow j}^{h^\prime}
    \Bigr]
  \prod\limits_{k\in c(j)}
  \Bigl[
    q_{k \rightarrow j}^0 + \sum\limits_{h^{\prime\prime} \geq h+1}^{D}
    q_{k\rightarrow j}^{h^{\prime\prime}}
    \Bigr] \; .
\end{equation}

The free entropy contribution of vertex $j$ is defined as
\begin{equation}
\phi_j(x)
\equiv \frac{1}{x} \ln \Bigl[ \frac{Z(x)}{Z(x; G_{\backslash j})}
  \Bigr] \; ,
\end{equation}
where $Z(x; G_{\backslash j})$ denotes the partition function of the cavity
digraph $G_{\backslash j}$. Under the Bethe-Peierls approximation, the
explicit expression of $\phi_j(x)$ is
\begin{equation}
\hspace*{-2.0cm}
  \phi_j(x) =
  \frac{1}{x}  \ln
  \biggl(
  1 + e^{x w_j} \sum\limits_{h=1}^{D}
  \prod\limits_{l\in b(j)} q_{l\rightarrow j}^0
  \prod\limits_{i \in p(j)}
  \Bigl[
    \sum\limits_{h^\prime=0}^{h-1} q_{i\rightarrow j}^{h^\prime}
    \Bigr] 
  \prod\limits_{k\in c(j)}
  \Bigl[
    q_{k\rightarrow j}^0 + \sum\limits_{h^{\prime\prime} \geq h+1}^{D}
    q_{k\rightarrow j}^{h^{\prime\prime}}
    \Bigr]
  \biggr) \; .
  \label{eq:phiv}
\end{equation}
Consider a link $(i\rightarrow j)$  which indicates a single arc $[i, j]$
between vertices $i$ and $j$. The free entropy contribution of this link is 
defined as
\begin{equation}
\phi_{(i\rightarrow j)}(x)
\equiv \frac{1}{x} \ln \Bigl[ \frac{Z(x)}
  {Z\bigl(x; G_{\backslash [i,j]} \bigr)}
  \Bigr] \; ,
\end{equation}
where $Z\bigl(x; G_{\backslash [i,j]}\bigr)$ is the partition function of
the cavity graph $G_{\backslash [i,j]}$ obtained by removing the single arc
$[i, j]$ from $G$. Its explicit expression under the Bethe-Peierls
approximation is
\begin{equation}
  \phi_{(i\rightarrow j)}(x) =
  \frac{1}{x}  \ln
  \biggl( q_{j\rightarrow i}^0 + \sum\limits_{h_j=1}^{D} q_{j\rightarrow i}^{h_j}
  \Bigl[\sum\limits_{h_i=0}^{h_j-1} q_{i\rightarrow j}^{h_i}
    \Bigr]
  \biggr) \; .
  \label{eq:phiarc}
\end{equation}
Similarly, the free entropy contribution $\phi_{(i\Leftrightarrow j)}(x)$
of a link $(i\Leftarrow j)$ (due to the existence of arcs $[i, j]$ and
$[j, i]$ of opposite direction between vertices $i$ and $j$)
is expressed as
\begin{equation}
  \phi_{(i\Leftrightarrow j)}(x) = \frac{1}{x} \ln \Bigl( q_{i\rightarrow j}^{0} +
  q_{j\rightarrow i}^0 - q_{i\rightarrow j}^0 q_{j\rightarrow i}^0 \Bigr) \; .
\end{equation}

Under the Bethe-Peierls approximation,
the whole free entropy $\Phi(x)$ of digraph $G$ can be evaluated through
the following simple expression:
\begin{equation}
  \label{eq:Phi}
  \Phi(x) = \sum\limits_{j=1}^{N} \phi_j(x) -
  \sum\limits_{(i\rightarrow j)\in G}
  \phi_{(i\rightarrow j)}(x)
  -\sum\limits_{(i\Leftrightarrow j)\in G} \phi_{(i\Leftrightarrow j)}(x)  \; .
\end{equation}
The first term of the above expression is the sum of free entropy
contributions of all the vertices. Since each link $(i\rightarrow j)$
contributes to the free entropies of vertices $i$ and $j$, the free entropy
contribution of this link should be subtracted once
from the total free entropy. This explains the second term of
(\ref{eq:Phi}). Similarly the free entropy contribution of each
link $(k \Leftrightarrow l)$ should be subtracted once from the total
free entropy, leading to the third term of (\ref{eq:Phi}).

The free entropy density is computed from (\ref{eq:Phi})
as $\phi(x) = \frac{1}{N} \Phi(x)$. The entropy density $s$ of the
system is then calculated through
\begin{equation}
  s = - x^2 \frac{\partial \phi(x)}{\partial x} =
  x \bigl[ \phi(x) - \omega(x) \bigr] \; ,
\end{equation}
where $\omega(x)$ is the relative total weight of the occupied vertices:
\begin{equation}
  \label{eq:omega}
  \omega(x)
  \equiv \frac{1}{N} \sum\limits_{j=1}^{N} (1-q_j^0) w_j \; .
\end{equation}

To actually compute the free entropy density $\phi(x)$ and other
thermodynamic quantities, we need to compute the
  two cavity probability distributions $p_{j\rightarrow k}^{h_j}$ and 
  $p_{k\rightarrow j}^{h_k}$ associated with every link $(j\rightarrow k)$
  and also the two distributions $p_{k\rightarrow j}^{h_k}$ and
  $p_{j\rightarrow k}^{h_j}$ associated with every link $(j\Leftrightarrow k)$.
Under the same Bethe-Peierls approximation we can write down
the following self-consistent equations for two connected
vertices $j$ and $j^\prime$:
\begin{equation}
  \label{eq:qjjpmarginal}
  q_{j \rightarrow j^\prime}^{0}
  =
  \frac{1}{z_{j\rightarrow j^\prime}} \; , \quad \quad \quad
  q_{j \rightarrow j^\prime}^{h_j}  =
  \frac{a_{j\rightarrow j^\prime}(h_j)}{z_{j\rightarrow j^\prime}} \; ,
  \quad\quad (1 \leq h_j \leq D)
\end{equation}
where $z_{j\rightarrow j^\prime}$ and
$a_{j\rightarrow j^\prime}(h)$ are computed through the following
two expressions, respectively:
\begin{eqnarray}
  \hspace*{-2.0cm} 
  z_{j\rightarrow j^\prime}  = 
  1 + e^{x w_j} \sum\limits_{h=1}^{D}
  \prod\limits_{l\in b(j)\backslash j^\prime} q_{l\rightarrow j}^0
  \prod\limits_{i \in p(j)\backslash j^\prime}
  \Bigl[
    \sum\limits_{h^\prime=0}^{h-1} q_{i\rightarrow j}^{h^\prime}
    \Bigr] 
  \prod\limits_{k\in c(j)\backslash j^\prime}
  \Bigl[
    q_{k \rightarrow j}^0 + \sum\limits_{h^{\prime\prime} \geq h+1}^{D}
    q_{k\rightarrow j}^{h^{\prime\prime}}
    \Bigr]
  \; , \\
\hspace*{-2.0cm}
a_{j\rightarrow j^\prime}(h)  = 
  e^{x w_j}
  \prod\limits_{l\in b(j)\backslash j^\prime} q_{l\rightarrow j}^0
  \prod\limits_{i \in p(j)\backslash j^\prime}
  \Bigl[
    \sum\limits_{h^\prime=0}^{h-1} q_{i\rightarrow j}^{h^\prime}
    \Bigr]
  \prod\limits_{k\in c(j)\backslash j^\prime}
  \Bigl[
    q_{k\rightarrow j}^0 + \sum\limits_{h^{\prime\prime} \geq h+1}^{D}
    q_{k\rightarrow j}^{h^{\prime\prime}}
    \Bigr] \; .
\end{eqnarray}
Notice that Eq.~(\ref{eq:qjjpmarginal}) holds for all vertices
$j^\prime \in \partial j$, and it differs from Eq.~(\ref{eq:qjmarginal})
only by discarding vertex $j^\prime$ from the set $\partial j$.
Equation (\ref{eq:qjjpmarginal}) is referred to as the
belief propagation (BP) equation.
This equation together with
the free entropy expression (\ref{eq:Phi}) forms the RS
mean field theory of  model (\ref{eq:Zxd1}).

The BP equation can be solved by iterations.
In each time step of this iteration we consider all the vertices
consecutively and in a random order, and for each vertex
$j$ under consideration we update all its output cavity
probabilities $q_{j\rightarrow j^\prime}^{h_i}$ using (\ref{eq:qjjpmarginal}).
The difference between the updated cavity
probability $q_{j\rightarrow j^\prime}^{h_i}(t+1)$ at time
$t+1$ and the old cavity probability $q_{j\rightarrow j^\prime}^{h_i}
(t)$ is measured by
\begin{equation}
\Delta_{j\rightarrow j^\prime}(t)  \equiv
\sum\limits_{h=0}^{D} \Bigl|
q_{j\rightarrow j^\prime}^{h}(t+1)-q_{j\rightarrow j^\prime}^{h}(t) \Bigr|
 \; .
\end{equation}
If the maximal value of all the differences
$\Delta_{j\rightarrow j^\prime}(t)$ is less than certain small threshold
$\Delta$ then we regard the BP iteration as converging
to a fixed point. In this work we set the convergence
criterion as $\Delta=0.01$.
Compared with the BP iteration process of the undirected
FVS problem \cite{Zhou-2013}, the BP iteration for
the directed FVS problem is more time- and computer
memory-consuming, since each height state can have $(D+1)$ possible values.

After the BP equation converges on all the arcs we can then compute the
free entropy density $\phi$ and the entropy density $s$. In addition, 
the mean fraction $\rho$ of vertices in the FVS is equal to the mean fraction
of unoccupied vertices:
\begin{equation}
  \label{eq:rhoexp}
  \rho =\frac{1}{N} \sum\limits_{j=1}^{N} q_j^0 \; .
\end{equation}

\subsection{Application to a single directed cycle}
\label{sec:cycle}

\begin{figure}
  \begin{center}
    \subfigure[]{
      \label{fig:cycle:a}
      \includegraphics[height=0.3\textwidth]{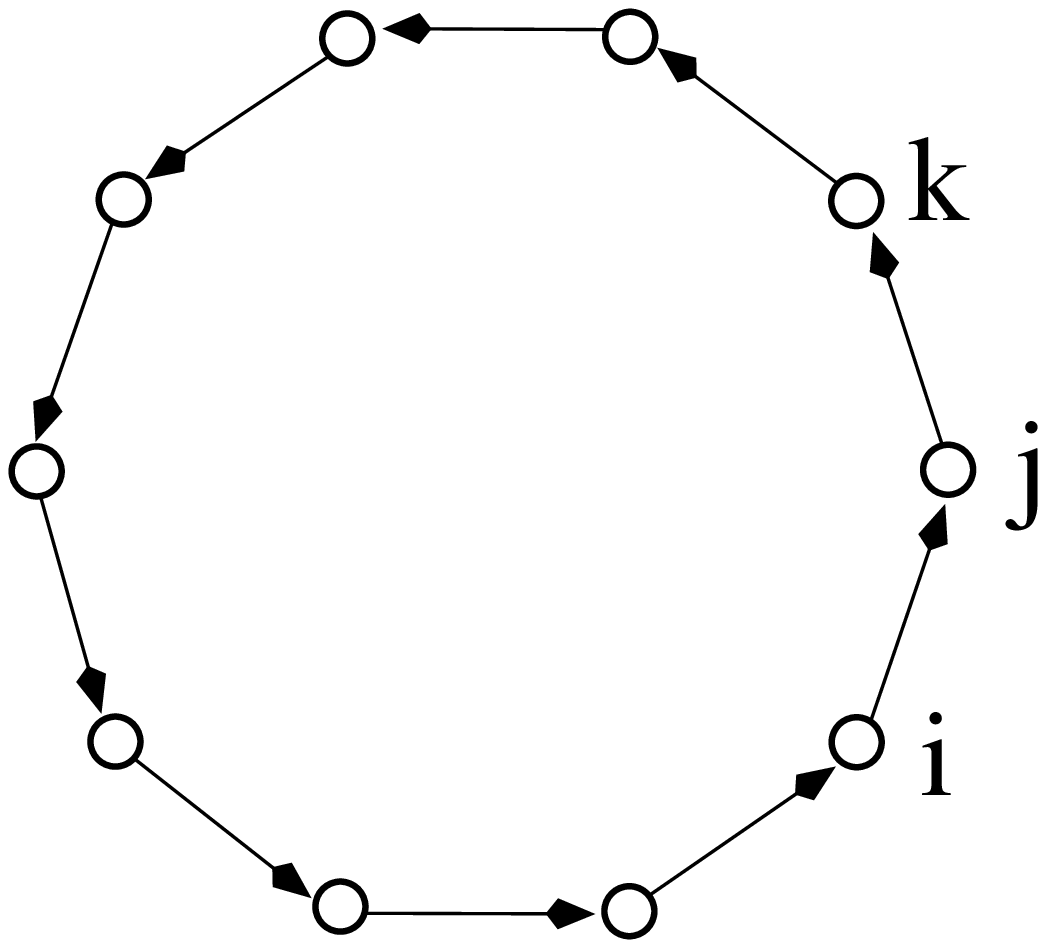}
    }
    \hskip 1.0cm
    \subfigure[]{
      \label{fig:cycle:b}
      \includegraphics[width=0.45\textwidth]{CycleResult.eps}
    }
  \end{center}
  \caption{\label{fig:cycle}
    Replica-symmetric mean field results for a single directed cycle.
    (a) A directed cycle with $N=10$ vertices. 
    (b) The entropy density $s$ versus
    FVS relative size $\rho$ obtained at three
    maximum height values $D=5$, $D=10$, and $D=15$.
  }
\end{figure}

As a simple illustration of the RS mean field theory, let us apply it to
a digraph composed of a single directed cycle of $N$ vertices and $M=N$
arcs (in the example of Fig.~\ref{fig:cycle:a}, $N=10$). The BP
equation (\ref{eq:qjjpmarginal}) always converges for this system, and
at the BP fixed point all the parent-to-child cavity distributions
$q_{i\rightarrow j}^{h}$ for arcs $[i, j]$ are identical to the same probability
function $q_{p c}^h$ which satisfies
\begin{equation}
 \hspace*{-2.0cm}
 q_{p c}^{0}  = 
  \frac{1}{1+ e^{x} \sum\limits_{h=1}^{D}
    \sum\limits_{h^\prime=0}^{h-1} q_{p c}^{h^\prime}
    }\; , \quad \quad \quad
  q_{p c}^{h}  = 
  \frac{e^{x} \sum\limits_{h^\prime=0}^{h-1} q_{p c}^{h^\prime}}
       {1+ e^{x} \sum\limits_{h=1}^{D}
         \sum\limits_{h^\prime=0}^{h-1} q_{p c}^{h^\prime}
       } \quad (h=1,2, \ldots, D) \; .
       \label{eq:cyclefix}
\end{equation}

The unique solution of this equation is
\begin{equation}
\hspace*{-1.0cm}
q_{p c}^0 = \frac{1}{A} \; , \quad \quad \quad 
q_{p c}^{h} = \frac{e^x}{A (A+e^x)} \bigl(1 +
  \frac{e^x}{A}\bigr)^{h} 
  \quad (h=1,2,\ldots, D)\; ,
\end{equation}
where the constant $A$ is determined by $A =  \bigl(1+\frac{e^x}{A} \bigr)^D$.
Similarly all the child-to-parent cavity
distributions $q_{j\rightarrow i}^{h}$ on the arcs $[i, j]$ are identical to
the same probability function $q_{c p}^h$ with $q_{c p}^0 = \frac{1}{A}$
and $q_{c p}^h = \frac{e^x}{A (A+e^x)} \bigl(1+\frac{e^x}{A}\bigr)^{D+1-h}$
($1\leq h \leq D$).

The mean field theory then predicts that the FVS relative size $\rho$,
the free entropy density $\phi$, and the entropy density $s$ are
\begin{equation}
  \rho = \frac{1}{1+\frac{e^x D}{A+e^x}} \; ,
  \quad \quad 
\phi = \frac{1}{x} \ln A \; , \quad \quad
s = \ln A - x (1-\rho) \; ,
\end{equation}
which are all independent of the cycle length $N$. As $x\rightarrow \infty$
we have $s\rightarrow 0$ and $\rho \rightarrow \frac{1}{D+1}$, see
Fig.~\ref{fig:cycle:b}. The minimum FVS size $\rho_0$ predicted by the RS 
theory is $\rho_0=\frac{N}{D+1}$, which is higher than the true minimum
size $1$ if $D<(N-1)$ and is lower than the true value if $D$ is longer than
the cycle length $N$.

This exactly solvable example clearly demonstrates that the RS mean field 
theory is only an approximate theory.  A major shortcoming of the RS theory
is that it neglects all the cycle-caused long range correlations
and all the cycle corrections to the free entropy $\Phi(x)$ 
\cite{Chertkov-Chernyak-2006b,Zhou-Wang-2012,Xiao-Zhou-2011,Zhou-etal-2011},
even though the constraints in the directed FVS and FAS problems are
induced by directed cycles. 

  In  \ref{sec:appendix} we consider another simple random graph example
  to discuss more about the relationship between predicted minimum FVS
  relative size ($\rho_0$) and the parameter $D$. The numerical results shown
  in Fig.~\ref{fig:RRRa20} demonstrates that the predicted value of
  $\rho_0$ decreases considerably with $D$ for an infinitely large random
  digraph.

\subsection{Application to single random digraphs}

We now perform BP iterations on two types of random digraphs,
namely Erd\"os-R\'enyi (ER) digraphs and regular random (RR) digraphs
\cite{He-Liu-Wang-2009,Albert-Barabasi-2002}. We generate a random
digraph $G$ of $N$ vertices and $M = \alpha N$ arcs in two steps.
First we generate an undirected random simple graph $G^\prime$ of
$N$ vertices and $M$ undirected
edges. Then we turn each undirected edge of $G^\prime$ into a
directed edge (an arc) by assigning it a direction uniformly at random
from the two possible directions. The resulting digraph is then the
digraph $G$. An undirected ER graph $G^\prime$ is generated from an empty
graph of $N$ vertices. Two different vertices $i$ and $j$
are randomly drawn from the whole set of $N$ vertices, if these two vertices
are not yet connected we draw an edge between them. This
edge addition process continues until $M$ edges have been created.
An undirected RR graph has the property that each vertex is attached by
exactly $d = 2 \alpha$ edges, therefore to generate such a graph we first
assign $d$ $`$half-edges' to each vertex, and then repeatedly glue two
randomly chosen half-edges into a full edge if this full edge is neither
a self-connection nor a multi-edge between two vertices.

\begin{figure}
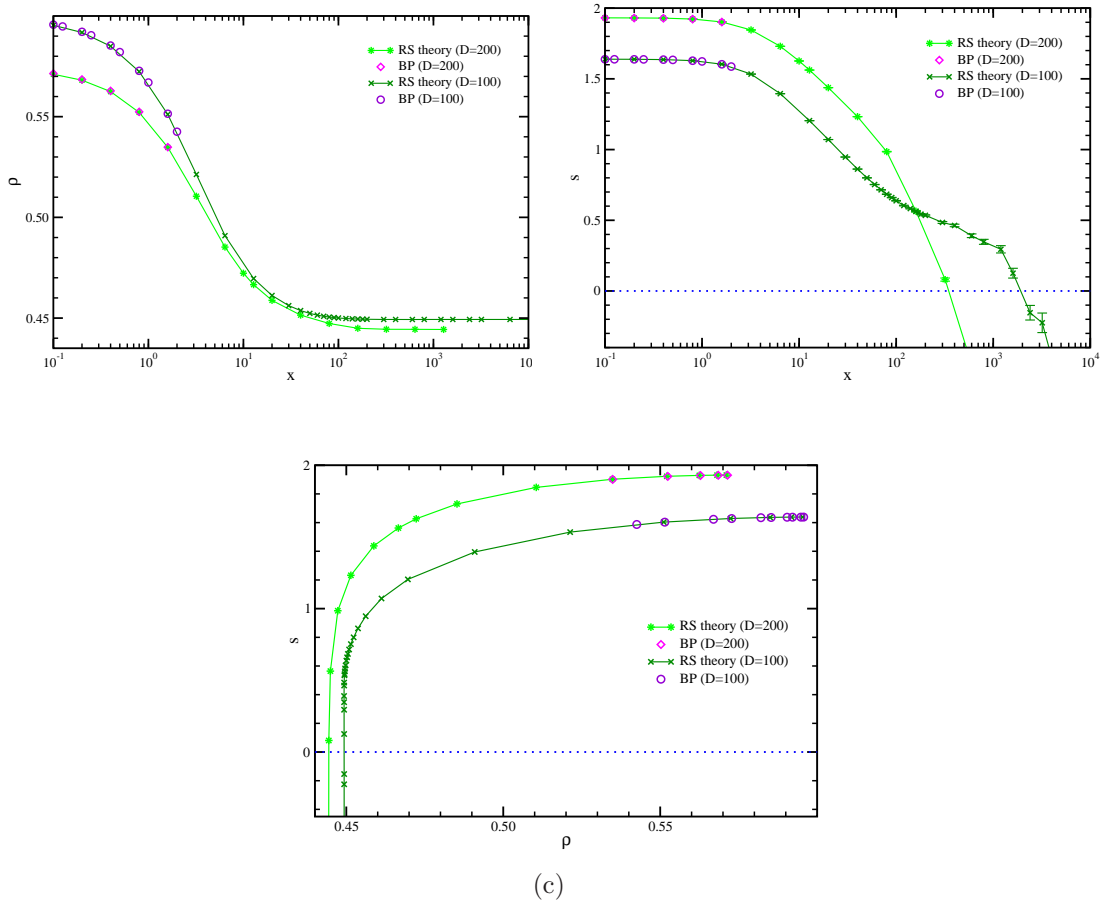

  \begin{center}
    \subfigure[]{
      \label{fig:EntropyRSpop10:a}
      \includegraphics[width=0.45\textwidth]{SDERa10FVS.eps}
    }
    \subfigure[]{
      \label{fig:EntropyRSpop10:b}
      \includegraphics[width=0.45\textwidth]{SDERa10Entropy.eps}
    }
  
    \subfigure[]{
      \label{fig:EntropyRSpop10:c}
      \includegraphics[width=0.45\textwidth]{SDERa10RS.eps}
    }
  \end{center}
  \caption{\label{fig:EntropyRSpop10}
    Replica-symmetric mean field results on
    ER random graphs of arc density $\alpha=10$
    and uniform vertex weight $w=1$, obtained with
    maximal height $D=100$ and $D=200$.
    (a) Mean fraction $\rho$ of unoccupied vertices. 
    (b) Entropy density $s$.
    (c) Entropy density $s$ as a function of FVS relative size
    $\rho$.
    The dotted lines of (b) and (c) indicate $s=0$.
    The BP results are obtained on a single digraph with $N=10^5$ vertices.
      The RS results are obtained by population dynamics simulations, which
      correspond to $N=\infty$.
  }
\end{figure}

The circular and diamond points of Figure~\ref{fig:EntropyRSpop10}
are the BP iteration results on a single random ER digraph with
$N=10^5$ vertices and arc density $\alpha = 10$. As the re-weighting 
parameter $x$ increases from zero, the relative size $\rho$ of FVS
and the entropy density $s$ decrease continuously.
However we find that the BP iteration process no longer converges
if $x$ exceeds $2$. The non-convergence of BP iteration
indicates the Bethe-Peierls approximation
is no longer a good approximation at $x > 2$ for this system (see also
Sec.~\ref{sec:cycle}).
On the other hand, when BP converges on a single digraph instance, the 
computed entropy density $s$ and FVS relative size $\rho$ are in good
agreement with ensemble-averaged results (see next subsection).

\subsection{Ensemble-averaged results}

The RS mean field theory can also be used to calculate
ensemble-averaged properties of random digraphs at the
thermodynamic limit of $N\rightarrow \infty$.
From Eq.~(\ref{eq:Phi}) we know that to
compute the free entropy density $\phi(x)$ we need only to
compute the mean value of vertex contribution
$\phi_j(x)$ over all vertices $i$
and the mean value of
link contribution $\phi_{(i\rightarrow j)}(x)$ over
all links $(i\rightarrow j)$ (the sum of
link contributions $\phi_{(i\Leftrightarrow j)}(x)$ can be
neglected since there is only a vanishing fraction of
links $(i\Leftrightarrow j)$ in random ER or RR digraphs).
These  mean values and the FVS relative size $\rho$ are easy to compute
by population dynamics simulations \cite{Mezard-Montanari-2009}.

Corresponding to Eq.~(\ref{eq:rhoexp}) for the FVS relative size $\rho$
of a single digraph, the ensemble-averaged value of $\rho$
is obtained by the following probabilistic equation (assuming all the
vertex weights $w=1$)
\begin{eqnarray}
  \hspace*{-1.5cm}
  \rho = \sum\limits_{d_{p}} \sum\limits_{d_c} P(d_{p}, d_{c}) 
  \prod\limits_{i}^{d_{p}}
  \int \mathcal{D} q_{i\rightarrow j} Q^{(p)}[q_{i\rightarrow j}]
  \prod\limits_{k}^{d_{c}}
  \int \mathcal{D} q_{k\rightarrow j}
  Q^{(c)}[q_{k\rightarrow j}] \nonumber \\
  \quad \quad \quad \quad \times
  \frac{1}{1 + e^{x} \sum\limits_{h=1}^{D}
    \prod\limits_{i \in p(j)}
    \Bigl[
      \sum\limits_{h^\prime=0}^{h-1} q_{i\rightarrow j}^{h^\prime}
      \Bigr]
    \prod\limits_{k\in c(j)}
    \Bigl[
      q_{k \rightarrow j}^0 + \sum\limits_{h^{\prime\prime} \geq h+1}^{D}
      q_{k\rightarrow j}^{h^{\prime\prime}}
      \Bigr]
  } \; .
  \label{eq:rhopd}
\end{eqnarray}
In this equation, $P(d_p, d_c)$ is the probability that a randomly chosen
vertex has $d_p$ parent vertices and $d_c$ child vertices. 
For ER digraphs $P(d_p, d_c)$ is the product of two
Poisson distributions with mean value $\alpha$,
$P(d_p, d_c)=\frac{e^{-\alpha} d_p^{\alpha}}{d_p!} 
\frac{e^{-\alpha} d_c^\alpha}{d_c!}$, 
while for RR digraphs $P(d_p, d_c)$ is the binomial distribution
$P(d_p, d_c)=\frac{(2 \alpha)!}{2^{2 \alpha}d_p! d_c!}\delta_{d_p+d_c}^{2 \alpha}$.
The probability functional $Q^{(p)}[q_{i\rightarrow j}]$ of 
Eq.~(\ref{eq:rhopd}) is the probability that the parent-to-child message on
an arc $[i,j]$ is the height distribution function
$q_{i\rightarrow j}^{h_i}$; similarly, $Q^{(c)}[q_{k\rightarrow j}]$ is the
probability that the child-to-parent message on an arc $[j,k]$ is
the height distribution function $q_{k\rightarrow j}^{h_k}$.

The ensemble-averaged expressions for
the vertex free-entropy contribution $\phi_i$ [Eq.~(\ref{eq:phiv})] is very
similar to Eq.~(\ref{eq:rhopd}), while the ensemble-averaged
expression for the arc free-entropy contribution $\phi_{(i\rightarrow j)}$
is computed through
\begin{equation}
  \frac{1}{x} 
  \int \mathcal{D} q_{i\rightarrow j} Q^{(p)}[q_{i\rightarrow j}]
  \int \mathcal{D} q_{j\rightarrow i}
  Q^{(c)}[q_{j\rightarrow i}] 
  \ln \biggl( q_{j\rightarrow i}^0 + \sum\limits_{h_j=1}^{D} 
  q_{j\rightarrow i}^{h_j}
  \Bigl[\sum\limits_{h_i=0}^{h_j-1} q_{i\rightarrow j}^{h_i}
    \Bigr]
  \biggr) 
  \; .
\end{equation}

In the population dynamics simulations the probability functional
$Q^{(p)}[q]$ is represented by a large array of $\mathcal{M}$ cavity
probability distributions $q_{j\rightarrow j^\prime}^{h_j}$ of the
parent-to-child
type (arcs $[j,j^\prime]$). Similarly $Q^{(c)}[q]$ is also represented by a
large array of $\mathcal{M}$ cavity probability distributions
$q_{j\rightarrow j^\prime}^{h_j}$ of the child-to-parent type
(arcs $[j^\prime,j]$).
To update an element of the array $Q^{(p)}[q_{j\rightarrow j^\prime}]$, we first
generate an in-degree $d_j^{(p)}$ and an out-degree
$d_j^{(c)}$ for the parent vertex $j$ according to the cavity degree 
distribution 
\begin{equation}
  \label{eq:cvPp}
  P^{(p)}(d_j^{(p)}, d_j^{(c)})
  \equiv \frac{ d_j^{(c)} P(d_j^{(p)}, d_j^{(c)})}{\alpha} \; ,
\end{equation}
and we then select $d_j^{(p)}$ parent-to-child probability distributions
$q_{i\rightarrow j}^{h_i}$ from the array $Q^{(p)}[q]$ and
$(d_j^{(c)}-1)$ child-to-parent probability distributions
$q_{k\rightarrow j}^{h_k}$ from the array $Q^{(c)}[q]$,
finally we compute a new probability distribution
$q_{j\rightarrow j^\prime}^{h_j}$ according to Eq.~(\ref{eq:qjjpmarginal}) and
replace an old element of the array $Q^{(p)}[q]$ with this new element.
The array $Q^{(c)}[q_{j\rightarrow j^\prime}]$ is updated according to the same
procedure, but the degree distribution of the child vertex $j$ is
changed to be
\begin{equation}
  \label{eq:cvPc}
  P^{(c)}(d_j^{(p)}, d_j^{(c)})
  \equiv \frac{ d_j^{(p)} P(d_j^{(p)}, d_j^{(c)})}{\alpha} \; .
\end{equation}

As an example, Fig.~\ref{fig:EntropyRSpop10} shows the ensemble-averaged 
mean field results for  ER digraphs with arc density
$\alpha=10$. In our calculations the weight of each vertex $i$ is set to
be $w_i=1$ for simplicity. The FVS relative size $\rho$ and the
entropy density $s$ both decrease with the re-weighting parameter $x$,
while the entropy density $s$ as a function of $\rho$ appears to be concave
and its value becomes negative as $\rho$ is lower than certain
threshold value $\rho_0(D)$ which depends slightly on the parameter $D$
(the maximal height), e.g., $\rho_0=0.449$ at $D=100$ and 
$\rho_0 = 0.444$ at $D=200$.  

\begin{figure}
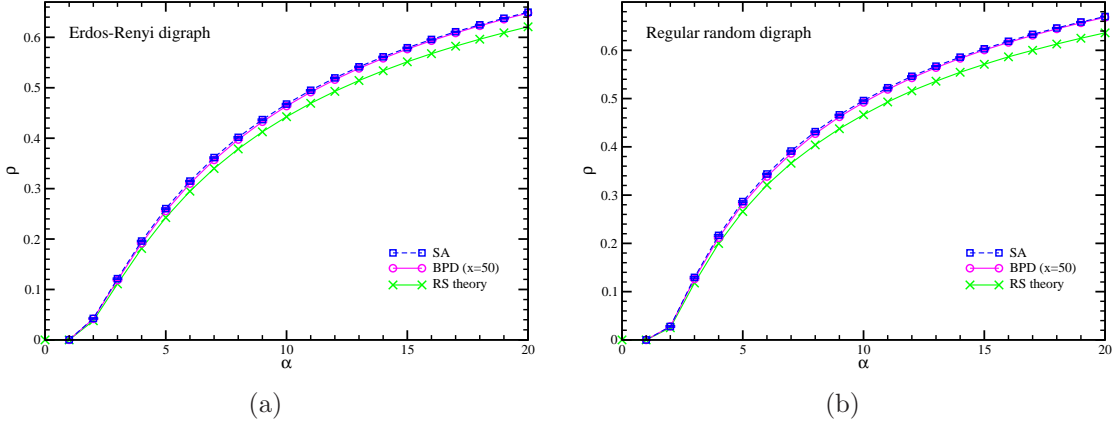

  \begin{center}
    \subfigure[]{
      \label{fig:ERfvs}
      \includegraphics[width=0.45\textwidth]{ERdfvs.eps}
    }
    \hskip 0.2cm
    \subfigure[]{
      \label{fig:RRfvs}
      \includegraphics[width=0.45\textwidth]{RRdfvs.eps}
    }
  \end{center}
  \caption{\label{fig:Rdfvs}
    Comparing the theoretical prediction and algorithmic results
    on the FVS relative size of ER (a) and RR (b) digraphs.
    Cross symbols are the RS mean field
    predictions on the minimum FVS relative size $\rho$
    obtained at $D=200$.
    Circular symbols (together with error bars) are
    the average values of the FVS relative size obtained by a
    single run of the BPD algorithm (with $D=200$ and $x=50$) on
    $96$ digraph instances of $N=10^4$ vertices and $M= \alpha N$ arcs.
    Square symbols (together with error bars)
    are the average values of the FVS relative size obtained by
    a single run of the SA algorithm (with the same parameters of
    \cite{Galinier-Lemamou-Bouzidi-2013})
    on these $96$ digraph instances.
  }
\end{figure}

At a given value of $D$ we take the value of 
$\rho_0$ obtained by the RS population dynamics at entropy density $s=0$
as the predicted minimum relative size of FVS.
The relationship between $\rho_0$ and the arc density $\alpha$ obtained at
$D=200$ is shown in Fig.~\ref{fig:Rdfvs} for ER and RR random digraphs.
In view of the discussion made in Sec.~\ref{sec:cycle} the predicted 
$\rho_0$ should only be considered as an educated guess on the
true value of FVS relative size.

\section{Belief propagation-guided decimation}
\label{sec:BPD}

Through BP iteration we can obtain an estimate about
the height probability distribution $q_j^{h_j}$ of each vertex $j$,
see Eq.~(\ref{eq:qjmarginal}). This estimate might not be very
accurate if the underlying Bethe-Peierls approximation is not a good
approximation (especially when the BP iteration fails to converge),
however it still contains very useful information for constructing
a near-optimal feedback vertex set.
We now describe a belief propagation-guided decimation (BPD) algorithm
for solving the directed FVS problem (and also the FAS problem)
on single digraph instances.

\subsection{Description of the algorithm}

Initially all the vertices of the digraph $G$ are declared as
active, and the candidate FVS is initialized as empty, 
$\Gamma= \emptyset$.
The BP iteration then
runs on this digraph for $T_0$ time steps and the height probability
distribution $q_j^{h_j}$ of each vertex $j$ is computed by
Eq.~(\ref{eq:qjmarginal}). We set $T_0=500$ in this
paper. The BPD algorithm then repeatedly performing the following
fixing-and-updating procedure:

In the fixing stage, the active vertices $j$ are ranked in descending order
of their empty probability $q_j^0$, and the vertices at the top 
$p$ percent of this ranked sequence are all  added to the set
$\Gamma$ and declared as inactive. 
We set $p=0.005$ in this paper.
The digraph induced by the remaining active vertices is then simplified by 
repeatedly turning an active
vertex into inactive if this vertex has no active parent and
brother vertices, or no active child and brother vertices.

Then in the updating state, we run BP iteration on the
simplified digraph of active vertices for $T$ time
steps ($T=10$ in this paper).
Then the height probability distributions of all the active vertices
are computed by Eq.~(\ref{eq:qjmarginal}) again.

All the vertices in the original digraph $G$ will be turned into
inactive by repeating the fixing-and-updating process.
The resulting set $\Gamma$ forms a FVS of $G$. We then further polish
this set by examining, in a random order, each vertex of $\Gamma$
and deleting it from $\Gamma$ if the reduced set is still
a FVS.
The performance of the BPD algorithm also depends on the maximal height
$D$ and the re-weighting parameter $x$.
For a digraph with $N$ vertices and $M$ arcs, the required memory
space for running the BPD algorithm is of order
$D \times M$ while the time complexity is of order
$(T_0+ T/p) \times D \times M$. 
The  source code of the BPD algorithm is accessible
at the author's webpage ({\tt power.itp.ac.cn/$\sim$zhouhj/codes.html}).

\begin{figure}
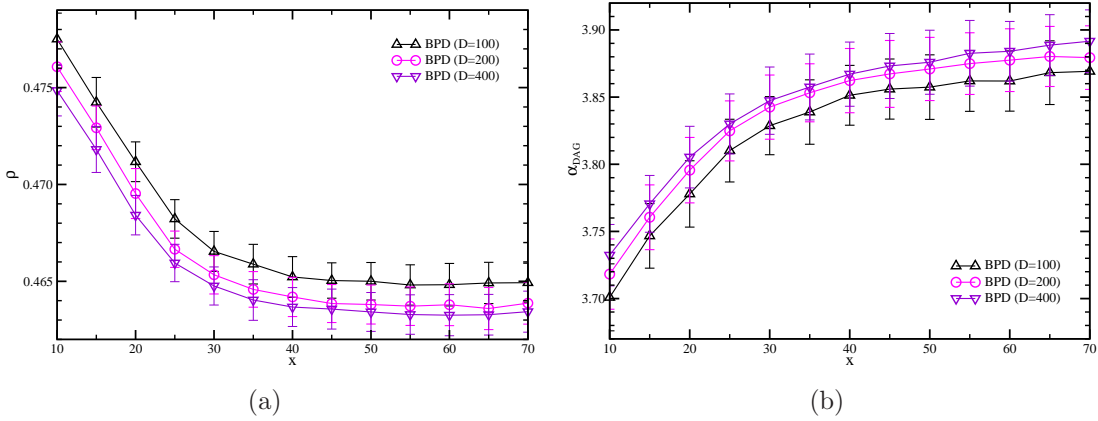

  \begin{center}
    \subfigure[]{
      \label{fig:BPD10:a}
      \includegraphics[width=0.45\textwidth]{BPDa10N10kFVS.eps}
    }
    \subfigure[]{
      \label{fig:BPD10:b}
      \includegraphics[width=0.45\textwidth]{BPDa10N10kEdgeDensity.eps}
    }
  \end{center}
    \caption{\label{fig:BPD10}
      The performance of the BPD algorithm on $96$ random ER digraphs of
      $N=10^4$ vertices and $M=10^5$ arcs 
      (arc density $\alpha = 10$). (a) Mean relative size $\rho$ of the
      constructed FVS.
      (b) Mean arc density $\alpha_{DAG}$ of the resulting directed acyclic
      graph.
      The maximal height is set to $D=100$ (up triangles),
      $D=200$ (circles), and $D=400$ (down triangles). The other parameters
      of the BPD algorithm are $T_0=500$, $T=10$, and $p=0.005$.
    }
\end{figure}

Figure~\ref{fig:BPD10} shows the mean value of the
relative sizes of constructed feedback vertex sets
and the mean value of arc densities of the complementary
directed acyclic digraphs, obtained by averaging the results of a
single run of the BPD algorithm on $96$ random ER digraphs of
arc density $\alpha =10$ and vertex number $N=10^4$. When the
maximum height $D$ is fixed, the performance of BPD improves
considerably as $x$ increases from $0$ to $30$ and then saturates
as $x$ goes beyond $40$. If the re-weighting parameter
$x$ is fixed but the maximal height changes from $D=100$ to $D=200$,
there is considerable improvement in the performance of BPD. However
if $D$ is further increased to $D=400$ the additional improvement in
performance is much weaker.
Based on the empirical observations of Fig.~\ref{fig:BPD10} we set
$x=50$ and $D=200$ when applying BPD to all 
the other random digraph instances mentioned below, even if these
digraphs have different arc density $\alpha$.

\subsection{Comparison with mean field predictions}

We apply the BPD algorithm to a set of ER digraphs with $N=10^4$ vertices
and mean arc density $\alpha$ ranging from $1$ to $20$. At each
value of $\alpha$ the FVS mean relative size, denoted as
$\rho_{BPD}$, is computed by averaging over 
the results of a single run of the
BPD algorithm on $96$ independent random digraph instances.
When $\alpha \leq 4$ we find the value of $\rho_{BPD}$
is very close to the minimum FVS relative size $\rho_{RS}$ as
predicted by the RS mean field theory at $D=200$,
see Fig.~\ref{fig:ERfvs}. 
However, the difference between $\rho_{BPD}$ and
$\rho_{RS}$ becomes noticeable at $\alpha \approx 5$ and the positive
gap $(\rho_{BPD} - \rho_{RS})$ increases continuously as $\alpha$
further increases.
  We have tested on several digraphs of larger size $N=10^5$ and found that
  the BPD results $\rho_{BPD}$ are not sensitive to $N$.
  At $D=200$, the discrepancy between the RS mean field results and
  the BPD algorithmic results therefore is unlikely to be caused
  by finite-size effects.
Instead we tend to believe that this gap indicates that the RS mean field
prediction $\rho_{RS}$ is lower than the true minimum
FVS size. We expect that the minimum FVS size obtained by the first-step 
replica-symmetric breaking  (1RSB) mean field theory
\cite{Mezard-Parisi-2001} will be closer to the empirical BPD results, but
we have not yet carried out such an investigation.
  We notice that an elegant 1RSB study has been carried out in the context
  of the minimal contagious set problem
  \cite{Guggiola-Semerjian-2015}, which is an inspiration for our future
  1RSB work.

We also apply the BPD algorithm to a set of RR digraphs with $N=10^4$
vertices and arc density $\alpha$ ranging from $1$ to $20$, and
compare the results with RS mean field predictions at $D=200$,
see Fig.~\ref{fig:RRfvs}.
When $\alpha \leq  3$ the mean FVS relative size
$\rho_{BPD}$ obtained by BPD 
is very close to the minimum FVS relative size $\rho_{RS}$ of
mean field theory, but the difference 
between  $\rho_{BPD}$ and
$\rho_{RS}$ becomes noticeable at $\alpha \approx 4$ and is more
and more pronounced as $\alpha$ further increases.

\subsection{Comparison with simulated annealing}

Recently an efficient heuristic local algorithm was proposed in
\cite{Galinier-Lemamou-Bouzidi-2013}, which repeatedly refine the height
configuration $\underline{h}$ of a digraph through simulated annealing
under the constraints (\ref{eq:linkc1}) and (\ref{eq:linkc2}). 
When tested on some small digraph instances with $N \leq 1000$ vertices,
the SA algorithm outperforms the greedy adaptive searching process (GRASP)
\cite{Pardalos-Qian-Resende-1999}, one of the most
successful local algorithms 
which  repeatedly reduces the number of directed cycles by deleting vertices
with highest value of in- and out-degree product $d_{p} \times d_{c}$.
At each elementary step of
the SA algorithm a trial is proposed to update the FVS:
(a) first a randomly chosen vertex $j$ of the FVS (height $h_j=0$) is chosen,
and (b) its height $h_j$ is either set to a maximal value while satisfying 
all the out-going arcs $[j, k]$ or set to a minimal value while satisfying
all the in-coming arcs $[i, j]$, and then (c)
the parent vertices $i$ of all the unsatisfied arcs $[i,j]$ 
or the child vertices $k$ of all the unsatisfied arcs $[j, k]$
are put to the FVS ($h_i=0$ or $h_k=0$).
Such a trial is accepted for sure if the resulting FVS size does not
increase, otherwise it is accepted with a low probability
\cite{Galinier-Lemamou-Bouzidi-2013}.

\begin{figure}
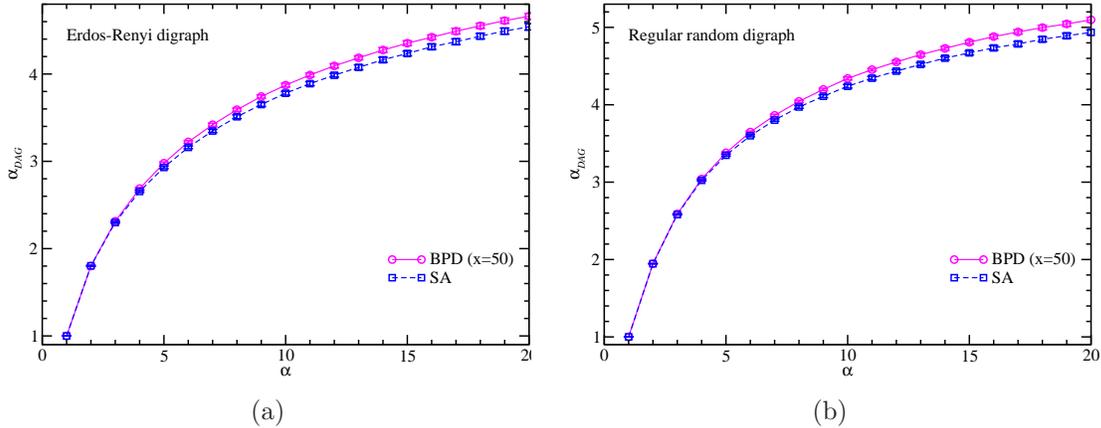

  \begin{center}
    \subfigure[]{
      \label{fig:ERdag}
      \includegraphics[width=0.45\textwidth]{ERdagDegree.eps}
    }
    \subfigure[]{
      \label{fig:RRdag}
      \includegraphics[width=0.45\textwidth]{RRdagDegree.eps}
    }
  \end{center}
  \caption{\label{fig:Rdag}
    The arc density $\alpha_{DAG}$ of the directed acyclic graph after
    all the vertices of the FVS and the attached arcs are removed from
    the input digraph of arc density $\alpha$.
    Circular symbols (together with error bars) denote the results obtained
    by the BPD algorithm ($D=200$ and $x=50$) on $96$ digraph instances of
    size $N=10^4$,
    while square symbols (together with error bars) denote the results 
    obtained by the SA algorithm (with the same parameters of
    \cite{Galinier-Lemamou-Bouzidi-2013}) on the same $96$ digraph instances.
    (a) ER digraph. (b) RR digraph.
  }
\end{figure}

Here we compare the performances of BPD and SA on large digraph instances
of $N=10^4$ vertices, see Fig.~\ref{fig:Rdfvs} and Fig.~\ref{fig:Rdag}.
First, we notice that 
the mean relative sizes of FVS constructed by BPD and by SA are
very close to each other both for ER and for RR digraphs. 
(We have also tested the BPD algorithm on the same set of small digraph 
instances used in \cite{Galinier-Lemamou-Bouzidi-2013} and found that BPD
performs equally good as SA.)
If we assume that the FVS solutions reached by the SA algorithm are close to
optimal solutions, the results of Fig.~\ref{fig:Rdfvs} then indicate that
(1) the BPD algorithm is able to construct close-to-minimum feedback vertex
sets, and that (2) the RS population dynamics predictions
underestimate the minimum FVS size of finite random digraph instances.

Although the sizes of FVS solutions constructed by BPD and by SA are almost
equal, we find that the BPD algorithm is more likely to select a vertex of
low connectivity into the FVS than the SA algorithm. As a result the DAG
obtained by the BPD algorithm has considerably higher arc density
(Fig.~\ref{fig:Rdag}). In other words, if we choose the BPD algorithm instead
of the SA algorithm, we can make the digraph free of directed cycles by 
deleting much fewer arcs. In this latter sense the BPD algorithm outperforms
the SA algorithm (besides the advantage that BPD is much faster
than SA).
  Of course both the BPD algorithm and the SA algorithm can be modified
  by adding another re-weighting parameter to avoid deleting highly connected
  vertices. Such an extension might be necessary for some practical 
  applications.

\section{Conclusion}
\label{sec:conclusion}

In this paper we introduced a spin glass model (\ref{eq:Zxd1}) for the 
directed feedback vertex set problem, approximately solved this model
by the replica-symmetric mean field theory of statistical physics,
and implemented a belief propagation-guided decimation algorithm to construct 
nearly optimal feedback vertex sets for single digraph instances.
The theory and algorithm of this paper is also applicable to the
equivalent feedback arc set problem.
The BPD algorithm slightly outperforms the simulated annealing algorithm
(Fig.~\ref{fig:Rdfvs} and Fig.~\ref{fig:Rdag}), it is therefore an efficient
algorithm to approach a minimum feedback vertex set for single difficult
digraph instances.
 
The RS mean field theory appears to compute a lower-bound for the FVS minimum
relative size $\rho$ (Fig.~\ref{fig:Rdfvs}). This theory assumes that the
height states of all the neighbors of a vertex $j$ are independent of each
other after vertex $j$ is deleted (Fig.~\ref{fig:BPapprox}). This approximation
essentially ignores all the correlations propagated along the directed 
cycles (e.g., in Fig.~\ref{fig:cycle:a} the heights $h_i$ and $h_k$ of 
vertices $i$ and $k$ must be strongly correlated at low temperatures even if
vertex $j$ is removed). Conceptually speaking, the RS mean field theory of
Sec.~\ref{sec:RS} can not be a satisfactory approach to tackle cycle
constraints.
  Much theoretical efforts (including 1RSB mean field computations)
  are needed to achieve a deeper understanding on the
  directed FVS problem.
An especially interesting but challenging question would be
to design a spin glass model with truly local interactions for the
cycle constraints (like the edge-constrained model for the undirected
FVS problem \cite{Zhou-2013}), without using height as the vertex state.

\section*{Acknowledgements}

The author thanks Yang-Yu Liu, Shao-Meng Qin,  Chuang Wang and
Jin-Hua Zhao for helpful discussions, and the School of Physics of
Northeastern Normal University for hospitality during his two visits in
August 2013 and August 2014. 
This work was supported by the National Basic Research Program of China
(grant number 2013CB932804), by the National Natural Science Foundation of
China (grant numbers 11121403 and 11225526), and by the
Knowledge Innovation Program of Chinese
Academy of Sciences (No.~KJCX2-EW-J02).


\begin{appendix}
 
  \section{A more restrictive spin glass model for the directed FVS problem}
  \label{sec:modelB}
  
  In this appendix we describe a more restrictive spin glass model for the
  directed feedback vertex set problem. The model in the main text can be
  regarded as a relaxed version of this model.
  For simplicity we assume that there is
  no bi-directional edges in the digraph $G$ so that the set of brother
  vertices of each vertex $j$ is empty: $b(j)= \emptyset$ . 
  The partition function of this new model is defined as
  \begin{equation} 
    \label{eq:modelB}
    Z(\beta) = \sum\limits_{\underline{h}}
    \prod\limits_{j=1}^{N} \Bigl[ e^{-x} \delta_{h_j}^0 +
      \delta_{h_j}^{1+max\{h_i : i\in p(j)\}} \Bigr] \; ,
  \end{equation}
  where $\max\{h_i : i \in p(j)\}$ returns the maximal height value among
  all the parent vertices of $j$.
  A height configuration $\underline{h} = (h_1, h_2, \ldots, h_N)$ with
  $N_0$ zero elements will contribute a term $e^{-x N_0}$ to this
  partition function if, for every vertex $j$, the height
  state $h_j$ is either zero (vertex $j$ being unoccupied) or $h_j$ is 
  exceeding the maximal height of its parent vertices by one.
  Such a height configuration is called a legal configuration, and all other
  height configurations are illegal and have no contribution to $Z(\beta)$.

  There is a one-to-one correspondence between a legal configuration
  $\underline{h}$ of model (\ref{eq:modelB}) and a FVS of the digraph $G$.
  First, it is obvious that the set formed by all the zero-height vertices of
  a legal configuration $\underline{h}$ must be a FVS. Second, given any
  feedback vertex set $\Gamma$ of the digraph we can construct a unique legal
  height configuration by recursion: first
  assign all vertices in $\Gamma$  the height value $0$ and put them into an
  initially empty vertex set $S$; then assign the height value $1$ to all the
  vertices whose parent vertices are completely contained in set $S$ and then
  add these newly assigned vertices to set $S$; then repeat this process and
  assign the remaining vertices the height values $2$, $3$, $\ldots$, until all
  the vertices have been exhausted. The resulting
  height configuration must be a legal configuration.
  
  The spin glass model (\ref{eq:modelB}) is more restrictive than the model
  (\ref{eq:Zxd1}). This is due to the fact that in the new model
  each vertex $j$ causes a 
  many-body constraint among $j$ and all its parent vertices. 
  For convenience of discussion let us denote
  by $\boxed{j}$ the constraint caused by vertex $j$.
  To solve this more difficult model by the replica-symmetric mean field
  theory, we denote by $p_{\boxed{\scriptstyle{k}}\rightarrow k}^{h_k}$ the
  probability that vertex $k$ will be in height state $h_k$ if it is only
  constrained by constraint $\boxed{k}$. Similarly, for each parent vertex $j$
  of vertex $k$, we denote by $p_{\boxed{\scriptstyle{k}}\rightarrow j}^{h_j}$ the
  probability that $j$ will be in height state $h_j$ if it is only constrained
  by the constraint $\boxed{k}$.
  We can write down the following set of belief-propagation equations
  for these two height distributions:
  \begin{eqnarray}
    & \hspace*{-2.2cm} p_{\boxed{{\scriptstyle{k}}}\rightarrow k}^{h_k} =
    \frac{1}{z_{{\boxed{{\scriptstyle{k}}}}\rightarrow k}}
    \biggl\{ e^{-x} \delta_{h_k}^0 + \delta_{h_k}^1
    \prod\limits_{j\in p(k)} q_{j\rightarrow \boxed{\scriptstyle{k}}}^{0}
    + \nonumber \\
    & \hspace*{2.0cm} (1- \delta_{h_k}^0 -\delta_{h_k}^1)
    \Bigl[\prod\limits_{j\in p(k)} Q_{j\rightarrow \boxed{\scriptstyle{k}}}^{h_k-1}
      -\prod\limits_{j\in p(k)} Q_{j\rightarrow \boxed{\scriptstyle{k}}}^{h_k-2} \Bigr]
    \biggr\} \; ,
    \label{eq:BPmany1}
    \\
    & \hspace*{-2.2cm} p_{\boxed{\scriptstyle{k}}\rightarrow j}^{h_j}  = 
    \frac{1}{z_{\boxed{\scriptstyle{k}}\rightarrow j}}
    \biggl\{ e^{-x} q_{k\rightarrow \boxed{\scriptstyle{k}}}^0 + 
    q_{k\rightarrow \boxed{\scriptstyle{k}}}^{h_j+1} \prod\limits_{i\in p(k)\backslash j}
    Q_{i\rightarrow \boxed{\scriptstyle{k}}}^{h_j} + \nonumber \\
    & \hspace*{2.0cm} \sum\limits_{h_k\geq h_j+2}
    q_{k\rightarrow \boxed{\scriptstyle{k}}}^{h_k} \Bigl[
      \prod\limits_{i\in p(k)\backslash j}
      Q_{i\rightarrow \boxed{\scriptstyle{k}}}^{h_k-1}
      -\prod\limits_{i\in p(k)\backslash j}
      Q_{i\rightarrow \boxed{\scriptstyle{k}}}^{h_k-2}
      \Bigr] \biggr\} \; .
    \label{eq:BPmany2}
  \end{eqnarray}
  In the above equations $z_{\boxed{\scriptstyle{k}}\rightarrow k}$ and
  $z_{\boxed{\scriptstyle{k}}\rightarrow j}$ are two probability normalization
  constants; the quantity
  $q_{k\rightarrow \boxed{\scriptstyle{k}}}^{h_k}$ denotes the probability that
  vertex $k$ will be in height state $h_k$ if it is \emph{not} constrained
  by constraint $\boxed{k}$; the quantity
  $Q_{j\rightarrow \boxed{\scriptstyle{k}}}^{h}$ is a partial sum defined as
  \begin{equation}
    Q_{j\rightarrow \boxed{\scriptstyle{k}}}^h \equiv
    \sum_{h_j=0}^h q_{j\rightarrow \boxed{\scriptstyle{k}}}^{h_j}
    \; ,
    \label{eq:BPmany3}
  \end{equation}
  where $q_{j\rightarrow \boxed{\scriptstyle{k}}}^{h_j}$ denotes the probability
  that the parent vertex $j$ of vertex $k$ will be in height state $h_j$
  if the constraint $\boxed{k}$ is absent.
  The self-consistent BP equations for
  $q_{k\rightarrow \boxed{\scriptstyle{k}}}^{h_k}$
  and $q_{j\rightarrow \boxed{\scriptstyle{k}}}^{h_j}$ are much simplier:
  \begin{eqnarray}
    q_{k\rightarrow \boxed{\scriptstyle{k}}}^{h_k} &=& 
    \frac{1}{z_{k\rightarrow \boxed{\scriptstyle{k}}}} 
    \prod\limits_{l\in c(k)} p_{\boxed{\scriptstyle{l}}\rightarrow k}^{h_k}
    \; ,
    \label{eq:BPmany4}
    \\
    q_{j\rightarrow \boxed{\scriptstyle{k}}}^{h_j}
    & = & \frac{1}{z_{j\rightarrow \boxed{\scriptstyle{k}}}} 
    p_{\boxed{\scriptstyle{j}}\rightarrow j}^{h_i} \prod\limits_{l\in c(j)\backslash k}
    p_{\boxed{\scriptstyle{l}}\rightarrow j}^{h_j}
    \; ,
    \label{eq:BPmany5}
  \end{eqnarray}
  where again $z_{k \rightarrow \boxed{\scriptstyle{k}}}$ and
  $z_{j\rightarrow \boxed{\scriptstyle{k}}}$ are  two probability normalization
  constants.
  
  The marginal probability $q_{j}^{h_j}$ that vertex $j$ will be in height
  state $h_j$ in the digraph $G$ is then evaluated at a fixed-point of the BP 
  iteration as
  \begin{equation}
    q_{j}^{h_j} = \frac{1}{z_j}
    p_{\boxed{\scriptstyle{j}}\rightarrow j}^{h_j}
    \prod\limits_{k\in c(j)} p_{\boxed{\scriptstyle{k}}\rightarrow j}^{h_j}
    \; ,
    \end{equation}
  with $z_j$ being a normalization constant. The mean fraction of unoccupied
  vertices (i.e., the FVS relative size) $\rho$ is computed as
  \begin{equation}
    \rho= \frac{1}{N} \sum\limits_{j=1}^{N} q_j^0 \; .
  \end{equation}

  The BP equations (\ref{eq:BPmany1})--(\ref{eq:BPmany5}) are rather slow
  to iterate. Our preliminary numerical results indicated that the feedback
  vertex set solutions offered by this more complicated BP scheme are not 
  better than the solutions obtained by the relaxed BP scheme of the main text.
  In this work we therefore give up further exploration of the model
  (\ref{eq:modelB}) of many-body interactions.

  To be complete, here we also list the explicit expression for the total
  free entropy $\Phi(x)$ of the new model:
  \begin{equation}
    \Phi(x)  = \sum\limits_{k=1}^{N} \Bigl[ \phi_{\boxed{\scriptstyle{k}}}
      - d_{k}^{(c)} \phi_{k} \Bigr] \,
  \end{equation}
  where $d_{k}^{(c)} \equiv | c(k)|$ is the number of child
  vertices of vertex $k$ (the out-degree). In this expression,
  $\phi_{\boxed{\scriptstyle{k}}}$ is the free entropy
  contribution of constraint $\boxed{k}$ and all its involved
  vertices, and $\phi_{k}$ is the free entropy contribution of vertex $k$.
  Their respective expressions are 
  \begin{eqnarray}
    & \hspace*{-1.0cm} 
    \phi_{\boxed{\scriptstyle{k}}}  = -\frac{1}{x} 
    \ln \biggl\{
    e^{-x} \prod\limits_{l\in c(k)} p_{\boxed{\scriptstyle{l}}\rightarrow k}^0
    \prod\limits_{j\in p(k)}\Bigl[
      \sum\limits_{h_j\geq 0}  p_{\boxed{\scriptstyle{j}}\rightarrow j}^{h_j}
      \prod\limits_{m\in c(j)\backslash k} 
      p_{\boxed{\scriptstyle{m}}\rightarrow j}^{h_j} \Bigr] \nonumber \\
    & \hspace{2.0cm}
    +  \prod\limits_{l\in c(k)} p_{\boxed{\scriptstyle{l}}\rightarrow k}^1
    \prod\limits_{j\in p(k)}\Bigl[
      p_{\boxed{\scriptstyle{j}}\rightarrow j}^{0}
      \prod\limits_{m\in c(j)\backslash k} 
      p_{\boxed{\scriptstyle{m}}\rightarrow j}^{0} \Bigr] \nonumber \\
    & \hspace{2.0cm} +
    \sum\limits_{h_k\geq 2}
    \prod\limits_{l\in c(k)} p_{\boxed{\scriptstyle{k}}\rightarrow k}^{h_k}
    \biggl(
    \prod\limits_{j\in p(k)}\Bigl[
      \sum\limits_{h_j=0}^{h_k-1} p_{\boxed{\scriptstyle{j}}\rightarrow j}^{h_j}
      \prod\limits_{m\in c(j)\backslash k} 
      p_{\boxed{\scriptstyle{m}}\rightarrow j}^{h_j} \Bigr] \nonumber \\
    & \hspace{4.0cm}
    - \prod\limits_{j\in p(k)}\Bigl[
      \sum\limits_{h_j=0}^{h_k-2} p_{\boxed{\scriptstyle{j}}\rightarrow j}^{h_j}
      \prod\limits_{m\in c(j)\backslash k} 
      p_{\boxed{\scriptstyle{m}}\rightarrow j}^{h_j} \Bigr] 
    \biggr)
    \biggr\}
    \; , \\
    & \hspace*{-1.0cm}
    \phi_{k}  = -\frac{1}{x} \ln \biggl\{
    \sum\limits_{h_k} p_{\boxed{\scriptstyle{k}}\rightarrow k}^{h_k} 
    \prod\limits_{l\in c(k)} p_{\boxed{\scriptstyle{l}}\rightarrow k}^{h_k}
    \biggr\}
    \; .
  \end{eqnarray}

  \section{Replica-symmetric solution for a special Random regular digraph}
  \label{sec:appendix}
  
  In Sec.~\ref{sec:cycle} we studied a single directed cycle and found that
  the RS mean field predicted that the relative size $\rho_0$ of minimum FVS
  approaches zero as the maximal height parameter $D\rightarrow \infty$. In
  this appendix we study another exactly solvable digraph and show that for
  this system the value of $\rho_0$ may converge to a strictly
  positive value at $D=\infty$.
  
  The digraph we now consider is a completely random digraph with the 
  constraint that each vertex has exactly $\alpha$ parent vertices and
  $\alpha$ child vertices (i.e., the in-degree and out-degree of every
  vertex is an integer $\alpha$). The weight of every arc in this digraph
  is set to be unity.
  For such a random digraph the BP equation
  (\ref{eq:qjjpmarginal}) has a fixed point with all the parent-to-child
  cavity distributions $q_{i\rightarrow j}^h$ for arcs $[i, j]$ being identical
  to the same probability function $q_{p c}^h$ and all the child-to-parent
  cavity distributions $q_{j\rightarrow i}^h$ being identical to the
  same probability function $q_{c p}^h$. Furthermore we find that
  $q_{c p}^0 = q_{p c}^0$ and $q_{c p}^h = q_{p c}^{D+1-h}$ for $1 \leq h \leq D$.
  
  Following the general BP equation (\ref{eq:qjjpmarginal}), we obtain
  the self-consistent condition for the function $q_{p c}^h$ as
  \begin{eqnarray}
    q_{p c}^0 & = &
    \frac{1}{1+e^x 
      \sum\limits_{\tilde{h}=1}^{D}
      \Bigl[ \sum\limits_{h^\prime=0}^{\tilde{h}-1} q_{p c}^{h^\prime} \Bigr]^{\alpha}
      \Bigl[ \sum\limits_{h^{\prime \prime}=0}^{D-\tilde{h}}
        q_{p c}^{h^{\prime \prime}} \Bigr]^{\alpha -1} } \; , 
    \label{eq:qpc0RRR} \\
    q_{p c}^h & = &
    \frac{e^x
      \Bigl[ \sum\limits_{h^\prime=0}^{h-1} q_{p c}^{h^\prime} \Bigr]^{\alpha}
      \Bigl[ \sum\limits_{h^{\prime \prime}=0}^{D-h}
        q_{p c}^{h^{\prime \prime}} \Bigr]^{\alpha -1} }
         {1+  e^x \sum\limits_{\tilde{h}=1}^{D}
           \Bigl[ \sum\limits_{h^\prime=0}^{\tilde{h}-1} q_{p c}^{h^\prime} 
             \Bigr]^{\alpha} \Bigl[ \sum\limits_{h^{\prime \prime}=0}^{D-\tilde{h}}
             q_{p c}^{h^{\prime \prime}} \Bigr]^{\alpha -1} } \quad \quad \quad
         (h=1, 2, \ldots, D) \; .
         \label{eq:qpchRRR}
  \end{eqnarray}
  These two equations have only a unique solution for any fixed values of
  the parameters $x$ and $D$. We determine the value of $q_{p c}^{h}$ through
  the following numerical procedure: 
  (1) set $q_{p c}^0$ to a value $q_0 \in (0,1]$; (2) determine the
    value $q_{p c}^D$ by solving 
    $q_{p c}^D = e^x q_0 q_0^{\alpha-1} (1-q_{p c}^D)^{\alpha}$, and then
    determine $q_{p c}^1$ according to
    $q_{p c}^1 = e^x q_0 q_0^{\alpha} (1- q_{p c}^D)^{\alpha - 1}$;
    (3) then determine the value of $q_{p c}^{D-1}$ by
    solving $q_{p c}^{D-1} = e^x q_0 (q_0 + q_1)^{\alpha-1} (1-q_{p c}^D
    -q_{p c}^{D-1})^{\alpha}$ and then determine $q_{p c}^2$ according to
    $q_{p c}^2=e^x q_0 (q_0+q_1)^{\alpha}(1-q_{p c}^D-q_{p c}^{D-1})^{\alpha-1}$;
    (4) then continue to determine all the remaining probability items
    $q_{p c}^{h}$ in the same sequential maner as steps (2) and (3);
    (5) then compute a new value of $q_{p c}^0$ according to
    Eq.~(\ref{eq:qpc0RRR}); (6) if $q_{p c}^0$ is different from the input
    value $q_0$ we then change $q_0$ appropriately and repeat steps
    (1)--(5) until convergence is reached.
    
\begin{figure}
  \begin{center}
    \includegraphics[angle=270,width=0.55\textwidth]{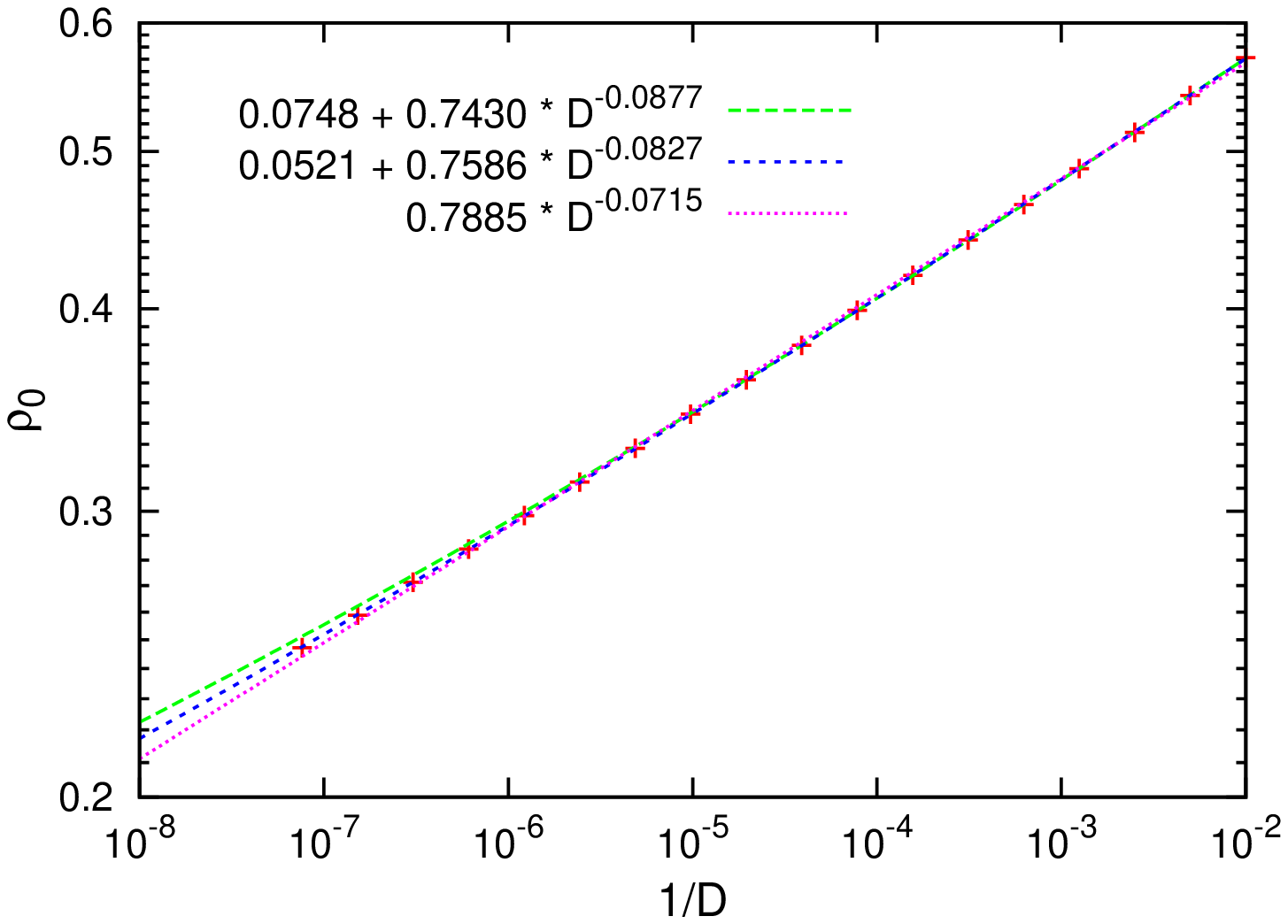}
  \end{center}
  \caption{\label{fig:RRRa20}
    The predicted minimum value $\rho_0$ of FVS relative size for the
    special regular random digraph in which every vertex is connected to
    $20$ parent vertices and $20$ child vertices. The value of $\rho_0$
    depends on the maximal height $D$ used in the RS mean field theory.
    The three fitting curves are $\rho_0 = 0.0748 + 0.7430 D^{- 0.0877}$
    (long-dashed blue line), $\rho_0 = 0.0521 + 0.7586 D^{-0.0827}$ (
    dashed blue line), and $\rho_0 = 0.7885 D^{-0.0715}$ (dotted red line).
  }
\end{figure}

    At the fixed point of Eqs.~(\ref{eq:qpc0RRR}) and (\ref{eq:qpchRRR}) we
    then compute the relative size $\rho$ of FVS is
    \begin{equation}
      \rho =  \frac{1}{1+e^x \sum\limits_{\tilde{h}=1}^{D}
        \Bigl[ \sum\limits_{h^\prime=0}^{\tilde{h}-1} q_{p c}^{h^\prime} \Bigr]^{\alpha}
        \Bigl[ \sum\limits_{h^{\prime \prime}=0}^{D-\tilde{h}}
          q_{p c}^{h^{\prime \prime}} \Bigr]^{\alpha -1} } \; .
    \end{equation}
    The free entropy density $\phi$ is
    \begin{eqnarray}
      \phi  &=& 
      \frac{\alpha}{x} \ln\biggl[
        1+  e^x \sum\limits_{\tilde{h}=1}^{D}
        \Bigl[ \sum\limits_{h^\prime=0}^{\tilde{h}-1} q_{p c}^{h^\prime} 
          \Bigr]^{\alpha} \Bigl[ \sum\limits_{h^{\prime \prime}=0}^{D-\tilde{h}}
          q_{p c}^{h^{\prime \prime}} \Bigr]^{\alpha -1} \biggr]  \nonumber \\
      & & 
      - \frac{(\alpha - 1)}{x} \ln \biggl[
        1+  e^x \sum\limits_{\tilde{h}=1}^{D}
        \Bigl[ \sum\limits_{h^\prime=0}^{\tilde{h}-1} q_{p c}^{h^\prime} 
          \Bigr]^{\alpha} \Bigl[ \sum\limits_{h^{\prime \prime}=0}^{D-\tilde{h}}
          q_{p c}^{h^{\prime \prime}} \Bigr]^{\alpha} \biggr] \; .
    \end{eqnarray}
    The entropy density is then simply computed as 
    $s= x \phi - x (1-\rho)$. At a given value of maximal height $D$ we
    can obtain the function $s(\rho)$ by performing numerical calculations
    at different values of $x$. The minimum relative size $\rho_0$ at this
    value of $D$ is then determined by solving $s(\rho_0)=0$. 
    The plus symbols of Fig.~\ref{fig:RRRa20} are the predicted values
    of $\rho_0$ up to $D=13,107,200$ for the particular case of arc density
    $\alpha=20$. Qualitatively the same mean field results are obtained for
    other values of the arc density $\alpha$.

    Figure~\ref{fig:RRRa20} demonstrates that $\rho_0$ decreases slowly with
    maximal height $D$. We can fit the behaviour of $\rho_0$ as
    \begin{equation}
      \label{eq:rho0fitting}
      \rho_0 = \rho_0^\infty + \frac{a}{D^\gamma} \; ,
    \end{equation}
    the fitting parameter $\rho_0^\infty$ is then the predicted minimum
    FVS relative size for an infinite random digraph. The fitted value of
    $\rho_0^\infty$ is sensitive to the range of $D$ values included in the
    fitting. For example, if we only use the data points of 
    $10^2 \leq  D\leq 10^5$, we
    obtain $\rho_0^\infty \approx 0.075 \pm 0.005$, see the green long-dashed 
    line of Fig.~\ref{fig:RRRa20}; 
    if we use the data points of $10^2 \leq D < 1.311 \times 10^7$ 
    we obtain a value $\rho_0^\infty \approx 0.052 \pm 0.002$,
    see the blue dashed line of Fig.~\ref{fig:RRRa20}.
    It's not easy for us to compute $\rho_0$ for 
    $D > 10^7$, but we anticipate that if more $\rho_0$ data obtained at
    very large $D$ values are included in the fitting, the fitted
    $\rho_0^\infty$ will further decrease. On the other hand, we observe
    that if we fix $\rho_0^\infty =0$ the fitting can not capture the 
    asymptotic behaviour of $\rho_0$ at $D> 10^6$ (see dotted red curve of
    Fig.~\ref{fig:RRRa20}). Therefore we believe $\rho_0^\infty$ should be
    strictly positive.

\end{appendix}

\section*{References}


\end{document}